\documentclass[journal]{IEEEtran}
\IEEEoverridecommandlockouts

\usepackage[utf8]{inputenc}
\usepackage[english]{babel}
\usepackage{graphicx}
\usepackage{xcolor}
\usepackage{cite}
\usepackage[T1]{fontenc} % optional enhanced font encoding
\usepackage{amsmath}
\usepackage{multirow}
\usepackage{bm}
\usepackage{array}
\usepackage{url}
\usepackage{psfrag}
\usepackage{mathtools}
\usepackage{amssymb}
\usepackage[linesnumbered,ruled]{algorithm2e}
\usepackage{algorithmic}

\usepackage{graphicx}
\usepackage{soul}
\usepackage[normalem]{ulem}
\usepackage[acronym]{glossaries}
\usepackage{caption}
\usepackage{subcaption}

\SetKwInput{KwInput}{Input}                % Set the Input
\SetKwInput{KwOutput}{Output}              % set the Output
\SetKwInput{kwInit}{Initialize}

\DeclareMathOperator*{\minimize}{minimize}

\begin{document}
%\captionsetup[figure]{name={Fig.},labelsep=period}
% acronym

% \newacronym[plural=cMs,firstplural=centiMorgans (cMs)]{cM}{cM}{centiMorgan}

\newacronym{wncs}{WNCS}{wireless networked control systems}
\newacronym{ul}{UL}{uplink}
\newacronym{dl}{DL}{downlink}
\newacronym{rx}{RX}{receiver}
\newacronym{tx}{TX}{transmitter}
\newacronym{awgn}{AWGN}{additive white gaussian noise}
\newacronym{snr}{SNR}{signal-to-noise ratio}
\newacronym{sdr}{SDR}{signal to distortion ratio}
\newacronym{mse}{MSE}{mean squared error}
\newacronym{gpr}{GPR}{gaussian process regression}
\newacronym{gp}{GP}{gaussian process}
\newacronym{lqr}{LQR}{linear quadratic regulator}
\newacronym{bs}{BS}{base station}
\newacronym{aoi}{AoI}{age-of-information}
\newacronym{dare}{DARE}{discrete-time algebraic Riccati equation}
\newacronym{iiot}{IIoT}{industrial internet of things}
\newacronym{csi}{CSI}{channel state information}
\newacronym{ml}{ML}{machine learning}
\newacronym{sc}{SC}{sensor-controller}
\newacronym{ca}{CA}{controller-actuator}
\newacronym{cdf}{CDF}{cumulative distribution function}
\newacronym{urllc}{URLLC}{ultra-reliable and low-latency communication}
\newacronym{rnn}{RNN}{recurrent neural network}
\newacronym{lstm}{LSTM}{long short-term memory}
\newacronym{nmpc}{NMPC}{nonlinear model predictive control}
\newacronym{ilqr}{iLQR}{iterative linear quadratic regulator}
\newacronym{rl}{RL}{reinforcement learning}
\newacronym{cococo}{CoCoCo}{communication and control co-design}
\newacronym{dmd}{DMD}{dynamic mode decomposition}
\newacronym{edmd}{EDMD}{extended dynamic mode decomposition}
\newacronym{ae}{AE}{autoencoder}
\newacronym{dkac}{DKAC}{deep koopman affine with control}
\newacronym{dkuc}{DKUC}{deep koopmanU with control}
\newacronym{sdp}{SDP}{semidefinite programming }
\newacronym{nn}{NN}{neural network}
\newacronym{mimo}{MIMO}{multiple-input multiple-output}
\newacronym{los}{LOS}{line of sight}
\newacronym{nlos}{NLOS}{none line of sight}
\newacronym{siso}{SISO}{single-input single-output}
\newacronym{mpc}{MPC}{model predictive control}
\newacronym{dnn}{DNN}{deep neural networks}

\title{Stability-Aware Joint Communication and Control for Nonlinear Control-Non-Affine Wireless Networked Control Systems}
\author{
\IEEEauthorblockN{
Rasika~Vijithasena\IEEEauthorrefmark{1}\IEEEauthorrefmark{2},~\IEEEmembership{Member,~IEEE,}
Rafaela~Scaciota\IEEEauthorrefmark{1}\IEEEauthorrefmark{2}, ~\IEEEmembership{Member,~IEEE,}
Mehdi Bennis\IEEEauthorrefmark{1},~\IEEEmembership{Fellow,~IEEE}, \\and 
Sumudu~Samarakoon\IEEEauthorrefmark{1}\IEEEauthorrefmark{2}~\IEEEmembership{Member,~IEEE}
} \\
\IEEEauthorblockA{
	\small%
	\IEEEauthorrefmark{1}%
	Centre for Wireless Communication, University of Oulu, Finland \\
    \IEEEauthorrefmark{2}
    Infotech Oulu, University of Oulu, Finland \\
	Email: \{rasika.vijithasena, rafaela.scaciotatimoesdasilva, mehdi.bennis, sumudu.samarakoon\}@oulu.fi %\\
	%\IEEEauthorrefmark{2}%
	%Institute 2, email: 
}
}

\maketitle

\begin{abstract}
Ensuring the stability of \gls{wncs} with nonlinear and control-non-affine dynamics, where system behavior is nonlinear with respect to both states and control decisions, poses a significant challenge, particularly under limited resources.  
However, it is essential in the context of $6$G, which is expected to support reliable communication to enable real-time autonomous systems.
This paper proposes a joint communication and control solution consisting of: i) a deep Koopman model capable of learning and mapping complex nonlinear dynamics into linear representations in an embedding space, predicting missing states, and planning control actions over a future time horizon; and ii) a scheduling algorithm that schedules sensor-controller communication based on Lyapunov optimization, which dynamically allocates communication resources based on system stability and available resources. 
Control actions are computed within this embedding space using a \gls{lqr} to ensure system stability.
The proposed model is evaluated under varying conditions and its performance is compared against two baseline models; 
one that assumes systems are control-affine, and another that assumes identical control actions in the embedding and original spaces.
The evaluation results demonstrate that the proposed model outperforms both baselines, by achieving stability while requiring fewer transmissions.
%--achieving stability
%-- fewer transmissions
\end{abstract}

\begin{IEEEkeywords}
Koopman operator, WNCS, Nonlinear, Machine learning, Deep Neural Network, Lyapunov Optimization
\end{IEEEkeywords}
\glsresetall
\section{Introduction} \label{sec:1}

The advancement of $5$G networks, alongside the emerging vision of $6$G technologies, is poised to significantly enhance the performance of \gls{wncs}. 
As key enablers to \gls{urllc}, \gls{wncs} play a vital role in achieving real-time operations and maintaining stability in dynamic, safety-critical applications~\cite{akyildiz20206g}. 
In contrast to traditional wired control systems, \gls{wncs} are closed-loop control systems where the communication between sensors, actuators and controllers takes place through wireless communication channels. 
This makes them particularly useful in situations where wired infrastructure is restrictive, costly, or impractical, such as in autonomous vehicles, \gls{iiot}, and remote monitoring systems~\cite{park2017wireless, liu2020latency}.

%Explaining problems
Despite these advancements, integrating wireless communication into control systems introduces several significant challenges. 
Wireless channels are inherently unreliable as they are influenced by varying environmental conditions such as multipath propagation, interference, fading, and shadowing, which lead to unpredictable latency and packet loss.
Resource constraints such as bandwidth and transmission power further exacerbate these issues~\cite{park2017wireless}. 
Additionally, many wireless nodes operate on limited battery power, restricting transmissions and power levels, which impacts \gls{urllc}, which require a significant amount of bandwidth, causing more transmission power~\cite{bennis2018ultrareliable}. 
These issues can severely impact the stability, controllability, and responsiveness of the control system, thereby necessitating advanced communication and control co-design strategies for the stringent performance requirements of next-generation \gls{wncs}~\cite{varghese2014wireless}.

%cococo
To address the challenge of limited resource allocation while ensuring system stability in \gls{wncs}, one common approach can be used is control-aware communication design, where communication policies are optimized under the assumption that control algorithms remain fixed, aiming to enhance the overall performance of the control system \cite{chang2019optimizing}. 
However, in practical \gls{wncs}, communication and control are often tightly coupled, resulting in significant interdependence between their respective policies~\cite{ ji2023intelligent}.
To more effectively manage this interdependence, \gls{cococo} has emerged as a widely adopted approach~\cite{9779178, fu2023smart, bhatia2021control}.
This approach jointly optimizes both communication and control policies to achieve a balance between communication constraints and control performance~\cite{pang2024communication}.
It ensures that control algorithms are robust to uncertainties in wireless communication, while communication strategies are tailored to meet the strict performance demands of control tasks by acknowledging the inherent interdependence between control and communication.

%with nonlinear systems
Furthermore, most of the existing research on \gls{cococo} has focused on linear control systems, leveraging their mathematical tractability to design predictive control strategies~\cite{9889798}. 
But in the real world, most control systems, such as robotic applications, chemical reactions, and autonomous vehicles, are often \textbf{nonlinear}, meaning their dynamics do not follow a linear relationship with the state of the system~\cite{adamy2022nonlinear}. 
In addition, they are frequently \textbf{control-non-affine}, which means that the relationship of control decisions with the dynamics of the system is not linear or directly separable.
This nonlinearity makes it significantly more difficult to accurately predict future system states and compute optimal control actions~\cite{sastry2013nonlinear}.

However, when extending these \gls{cococo} designs to nonlinear systems, it is possible to linearize the nonlinear dynamics around a local equilibrium point and approximate the behavior of the nonlinear system with a linear system in the neighborhood of that equilibrium point~\cite{tailor2011linearization, 9493202}.
%\textcolor{red}{In \cite{9493202}, authors have proposed a Predictive Control and Communication Co-Design which for achieving joint problem of uplink-downlink scheduling and power allocation for controlling multiple control systems but they have linearized the control system dynamics around equilibrium point.}
This approximation enables the use of classical linear control techniques, such as the \gls{mpc} and \gls{lqr}, to determine control actions~\cite{sastry2013nonlinear}. 
However, this approach becomes inaccurate when the system states deviate farther from the equilibrium, where the linear approximation is not applicable. 
Furthermore, this approach cannot be applied when system dynamics are unknown.
Moreover, these techniques are inapplicable when the system dynamics are unknown or difficult to model, highlighting the need for more robust approaches to control nonlinear \gls{wncs}.

Recent advances in \gls{ml}, particularly in time-series prediction models, have shown promise in addressing these challenges in accurately modeling and predicting the behavior of complex, real-world systems. 
\gls{rnn}, such as \gls{lstm} networks~\cite{bahdanau2014neural} and Bayesian models such as \gls{gpr}~\cite{liu2020gaussian} have been employed for predictive modeling.
While \gls{lstm} can capture temporal dependencies, its prediction accuracy degrades as the prediction horizon increases. 
\gls{gpr}, while effective for modeling uncertainty, suffers from computational inefficiencies and scalability issues in high-dimensional, nonlinear systems. 
Its performance is also highly sensitive to the choice of kernel functions.
Furthermore, after state prediction, for controlling nonlinear systems, nonlinear control methods such as \gls{nmpc}, \gls{ilqr}, and \gls{rl} are commonly employed \cite{redder2019deep}.
However, these \gls{nmpc} and \gls{ilqr} approaches often require extensive computational time, which limits their applicability in real-time, high-dimensional control systems.
\gls{rl} models require large datasets for training and have limitations in generalization and scalability when it comes to \gls{wncs} \cite{baumann2018deep, funk2021learning}.

Modern \gls{ml} research often takes inspiration from the human learning process, focusing on models that emulate how humans learn by building internal representations, planning actions, and predicting outcomes with minimal trial-and-error.
Inspired by this, the concept of "world models" has emerged in \gls{ml} research~\cite{vafa2024evaluating}. 
These models learn an internal representation of the system dynamics, which can be used for long-term state prediction and decision-making without requiring continuous access to real-time system states.
A promising framework in this context is the Koopman Operator theory~\cite{brunton2016koopman, koopman1932dynamical}, which enables transforming nonlinear dynamics into a representation space called the latent space or embedding space, where system's evolution can be approximated by linear dynamics, facilitating efficient linear control methods and long-horizon predictions even in high-dimensional nonlinear systems. 
This global linearization facilitates the use of linear control methods for calculating control actions in the embedding space and long-horizon predictions, even in high-dimensional nonlinear systems~\cite{kaiser2021data}.
Recent studies have explored data-driven approaches to model the Koopman operator for real-time nonlinear control, with \gls{dnn}-based methods proposed to jointly learn both the embedding functions and the Koopman operator in a fully data-driven manner.

%Recent studies have explored data-driven approaches that model the Koopman operator as a promising alternative for real-time control of nonlinear systems. 
%The Koopman operator provides a linear, but infinite-dimensional representation of nonlinear dynamics by evolving a set of embedding (observable) functions over time~\cite{brunton2016koopman}.
%\gls{dmd}~\cite{tu2013dynamic, proctor2018generalizing} can be used to approximate the Koopman operator, but a key challenge lies in DMD is based on linear measurements, which do not span a Koopman invariant subspace for many nonlinear systems~\cite{kaiser2021data}.
%\gls{edmd}~\cite{williams2015data} improves upon \gls{dmd} to approximate the Koopman operator using data-driven methods, but the limitation is a requirement of a predefined dictionary of functions of state, such as polynomials, trigonometric functions, or radial basis functions. 
%However, \gls{edmd}’s performance heavily depends on the choice of the dictionary; if the dictionary fails to capture the system’s nonlinear structure, the resulting Koopman approximation becomes inaccurate~\cite{hasnain2020steady}.

%literature review
In ~\cite{yeung2019learning} authors proposed a deep learning model for learning Koopman operators of nonlinear systems.
However, their approach focuses solely on multi-step prediction and does not consider controlling the system.
Extending this line of work, the authors of~\cite{shi2022deep} proposed an end-to-end deep learning framework to jointly learn the Koopman embedding function and Koopman operator, enabling state prediction for nonlinear systems and computing control decisions using \gls{lqr}. 
This framework includes an auxiliary control network to model state-dependent nonlinearities in the control input. 
While it performs well for control-affine cases, it lacks a conversion mechanism for applying control actions from the latent space to the real space in nonlinear control-non-affine scenarios, limiting its broader applicability.
Furthermore, in~\cite{vinod2024system}, authors proposed a Koopman-based approach for determining optimal control in nonlinear control systems. 
However, their method does not address the challenge of mapping control actions from the embedding space back to the original space, a problem that has been highlighted in prior literature.

In the context of \gls{wncs}, the work by~\cite{9889798} introduces a two-way Koopman autoencoder framework designed to improve state estimation and control in a wireless communication context. 
Specifically, this framework utilizes a sensing Koopman \gls{ae} to predict missing system states and a controlling Koopman \gls{ae} to predict missing control actions, corresponding to cases where communication has failed due to packet losses. 
A limitation of this approach is that the control system in which the predictions and control action calculations are made is assumed to be nonlinear and control-affine. 
Additionally, it is assumed that the control action in the embedding space is equivalent to the action in real space. 
These assumptions could lead to inaccuracies in prediction outcomes.
Therefore, a gap still exists in the literature regarding the prediction and control of nonlinear, control-non-affine \gls{wncs} systems without compromising generality. 

In this paper, we present a novel approach for stabilizing nonlinear, control-non-affine \gls{wncs} under limited communication resources. 
Toward this end, we propose a Deep Koopman operator-based solution that utilizes the world model concept to predict and plan control actions in embedding space, consisting of a scheduling algorithm to optimize limited resource utilization.
Specifically, we map original system states and control actions to separate linear embedding spaces under the assumption that they do not share mixed observables. That is, the embedding functions depend only on either the state or the control action, not on both.  
It enables us to transform the embedding space control actions to the original space (referred to as control action decoding), thus resolving a limitation in the prior research.
Then the joint control and communication in \gls{wncs} are carried out by three tasks:
\begin{enumerate}
    \item Sensor-to-controller communication scheduling, using Lyapunov optimization to minimize the number of transmissions and thereby optimize communication resource usage.
    \item Prediction of state in the embedding space using the Koopman operator, with control actions computed via \gls{lqr}.
    \item Planning future control actions for multiple consecutive timesteps and sending them to the actuator to cache when the communication is reliable, to handle possible losses in the controller-actuator communication link.
\end{enumerate}

The deep Koopman model is trained using unsupervised learning techniques, making it suitable for systems with nonlinear and control-non-affine dynamics. 
We validate our proposed method on two control systems: a double pendulum connected by a spring (representing a nonlinear, control-non-affine system) and the classic cart pole (representing a control-affine system). 
The proposed model's performance is evaluated under various conditions, including different outage probabilities and control design parameters, and compared against existing baseline models.

%consider on ml technologies for this
%Indeed, to address the challenges posed by wireless communication limitations and the complexity of nonlinear dynamics various studies have been carried out in literature. 

The paper is structured as follows. 
Section~\ref{sec:2} introduces the system model architecture.
Section~\ref{sec:3} formulates the joint communication and control problem, and Section~\ref{sec:4} presents the proposed solution.
In Section~\ref{sec:5}, numerical evaluations and results are provided.
Section~\ref{sec:6} concludes the paper.

\textit{\textbf{Notation}}: Scalars are represented by lowercase letters, while sets are indicated using calligraphic letters. 
Bold lowercase letters represent column vectors, and bold uppercase letters are used for matrices.
Real valued $D$ dimensional vector is denoted by $\mathbb{R}^D$.
Size of $D \times D$ real-valued matrix, complex valued matrix and positive semi-definite matrix are represented by $\mathbb{R}^{D \times D}, \mathbb{C}^{D \times D}$ and $\mathbb{S}^{D}$ respectively. 
Statistical expectation is denoted by $\mathbb{E}[.]$ and maximum value of two values is represented as $\max\{.,.\}$.
The operators $\|. \|$, $ (.)^T$ represents euclidean norm and trace respectively.
While the probability is represented as Pr(.) and the \gls{cdf} function of $h$ is denoted by $F_h(.)$.
A complex Gaussian distribution with mean $\mathbf{m}$ and covariance  $\mathbf{R}$ is denoted by $\mathcal{CN}(\mathbf{m}, \mathbf{R})$.

\section{System Model}\label{sec:2}

\begin{figure}
    \centering
    \includegraphics[width=1\linewidth]{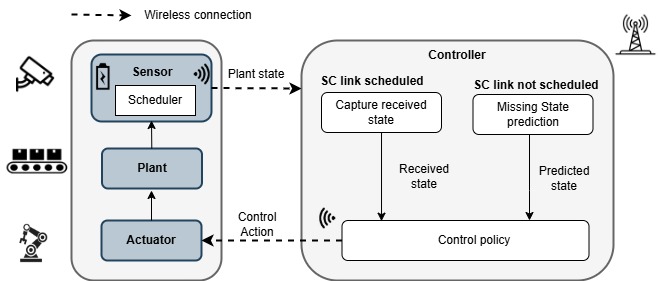}
    \caption{The system model of a \gls{wncs} setting illustrating a plant, sensor, actuator and controller where the controller calculates control actions using either predicted state or received state.}
    \label{fig:system_model}
\end{figure}

%\subsection{Control System}

We consider a non-linear, control-non-affine \gls{wncs} setting consisting of a plant, sensor, controller, and actuator as presented in Fig. \ref{fig:system_model}. 
The sensor and actuator are located in the plant, while the controller is located separately in a \gls{bs}.
Communication between the \gls{sc} and the \gls{ca} is conducted over wireless links.

The sensor captures the $D$-dimensional plant's state $\mathbf{x}_{t} = [{x}_{t}^d]_{d \in \{1,...,D\}}$ at each time $t$ and sends to the controller over a wireless connection.
Then, with the received sensor state, the controller computes and sends the $D'$-dimensional control action $\mathbf{u}_{t} \in \mathbb{R}^{D'} $ to the actuator through a wireless channel, which is then applied to the plant.
The physical plant is modeled as a discrete-time non-linear control-non-affine system given as~\cite{9889798},
\begin{equation}\label{eq:state_dynamic}
    \mathbf{x}_{t+1} = \mathrm{f}(\mathbf{x}_{t},\mathbf{u}_{t}) + \mathbf{n}_{t}
\end{equation}
where $\mathrm{f}\colon\mathbb{R}^D \times \mathbb{R}^{D'} \to \mathbb{R}^D$ is a non-linear function describing the state transition, and $\mathbf{n}_{t} \in \mathbb{R}^D$ denotes the $D$-dimensional plant noise assumed to be independently and identically distributed (i.i.d.) Gaussian with zero mean and the covariance matrix $\mathbf{N}$.
%that is sampled from an independent and identically distributed (i.i.d.) Gaussian distribution with zero mean and the covariance of $\mathbf{N}$.

The sensor is powered by a battery, with the available battery power at each time $t$ is denoted by $p_{\text{b},t}$.
Battery power is consumed for sensing the states of the plant and \gls{sc} communication.
Consequently, the battery power dynamics evolve by depleting the available battery power by $p_{\text{s},t}$ for sensing and $p_{\text{SC},t}$ for \gls{sc} transmission at each time $t$. 
We assume that the battery is initially fully charged with an initial battery power $p_{\text{b},0}$, and it is recharged every $T'$ time units.
Furthermore, the actuator includes a cache memory to store received control actions, allowing actuator to apply previously stored control action during \gls{ca} communication failures.

\subsection{Wireless Communication Model}

The system is assumed to be \gls{siso} for both the \gls{sc} and \gls{ca} links, where a single antenna is used at both the transmitter and the receiver.
The channels between the \gls{sc} and \gls{ca} links are subject to Rician fading with \gls{los} factor $\kappa_i$ where $i \in \{\text{SC}, \text{CA}\}$. 
Rician fading accounts for the presence of a dominant \gls{los} component, which is often observed in structured environments characterized by organized layouts and controlled object placement.
Such conditions are typical in scenarios where \gls{wncs} are deployed, such as \gls{iiot} settings and autonomous vehicle networks~\cite{abdi2002estimation}. 
The channel coefficient ${h}_{i,t}$ is defined as~\cite{proakis2008digital},
\begin{equation}
    {h}_{i,t} = \sqrt{\frac{\kappa_i}{1+\kappa_i}} \bar{{h}}_{i,t} + \sqrt{\frac{1}{1+\kappa_i}} \Tilde{{h}}_{i,t},
\end{equation}
where $\Tilde{{h}}_{i,t} \in \mathbb{C}$ is \gls{nlos} component modeled as a complex Gaussian variable with independent and identically distributed (i.i.d.) random entries drawn from a symmetric complex Gaussian distribution with zero mean and variance $\mathbf{\sigma}^2$ denotes as,  $ \mathcal{CN}(\mathbf{0},\mathbf{\sigma}^2)$. 
The LOS component $\bar{{h}}_{i,t} = e^{j \varphi}$ where $\varphi$ is the corresponding \gls{los} phase shift between transmitter and receiver antennas \cite{zhao2017capacity}.
The \gls{snr} is given by~\cite{9493202}
 \begin{equation} 
 \label{eq11}
 \gamma_{i,t} = \frac{p_{i,t} | {h}_{i,t} |^{2}}{N_{0}\omega },
 \end{equation}  
where $p_{i,t}$, $\omega$ and ${N_{0}}$ correspond to the transmission power, bandwidth and the noise power spectral density, respectively. 
A packet is successfully transmitted if the instantaneous \gls{snr} at the destination link $\gamma_{i,t}$, for $i \in \{\text{SC}, \text{CA}\}$, exceeds the \gls{snr} threshold $\gamma_{0} = 2^{R_b}-1$, corresponding to a spectral efficiency of $R_b$ bits/s/Hz under the Shannon limit. 
Thus, the outage probability can be written as~\cite{proakis2008digital}:
\begin{equation} 
 \label{eq.:outage}
\mathrm{O}_{i} = \Pr\left(\gamma_{i} < \gamma_{0}\right) = \Pr\left(| {h}_{i,t} |^{2} < \frac{\gamma_{0} N_{0}\omega}{p_{i,t}}\right).
 \end{equation} 

We define that channel is assumed to follow a known statistical distribution over time.
Consequently, the overall outage probability can be calculated as the average of the received power over the \gls{cdf} of $| {h}_{i,t} |^{2} $, such that~\cite{tellambura1997analysis}
\begin{align} \label{eq:outage_final}
    \mathrm{O}_{i} &= F_{| {h}_{i,t} |^{2}}\left(\frac{\gamma_{0} N_{0}\omega}{p_{i,t}}\right) \nonumber \\
    &= 1- Q_{1}\left(\sqrt{2\kappa_i} , \sqrt{2(1+\kappa_i)\frac{\gamma_{0} N_{0}\omega}{p_{i,t}}}\right),
\end{align}
where $Q_1(\cdot,\cdot)$ is the generalized Marcum $Q$-function of order $1$.
The transmission power $p_i$ in $i \in \{ \text{SC}, \text{CA} \}$ is chosen such that it meets the minimum required power to ensure $\mathrm{O}_i \leq \Tilde{\mathrm{O}}$, where $\Tilde{\mathrm{O}}$ is threshold for outage probability.

\subsection{Sensor-Controller Link Scheduling}

Due to the limited power availability at the sensor, it is assumed that the sensor employs a scheduler for the \gls{sc} transmission. 
Let $a_t$ be the indicator of \gls{sc} link scheduling defined as follows:
\begin{equation}\label{eq:scheduling_indicator}
a_t = \begin{cases} 
1 , & \text{if \gls{sc} link is scheduled}, \\
0 , & \text{otherwise}.
\end{cases}
\end{equation}
When $a_t = 1$, the system state is transmitted and because of that the controller can compute the control actions. 
In the absence of \gls{sc} communication, the controller predicts the system state, $\hat{\mathbf{x}}_t$, using past states, and calculates control actions. 
In that sense, we consider $\Tilde{\mathbf{x}}_t$ as the system state that the controller uses to compute control actions at each time $t$, 
\begin{equation}\label{eq:state_in_controller}
\Tilde{\mathbf{x}}_t = \begin{cases} 
\mathbf{x}_t, & \text{if $a_t = 1$}, \\
\hat{\mathbf{x}}_t = \hat{f}(\Tilde{\mathbf{x}}_{t-1}, \mathbf{u}_{t-1}) , & \text{otherwise},
\end{cases}
\end{equation}
where $\hat{f}(.)$ is a mapping function from the past state and past control action to the next state, possibly used in a recursive manner.

\subsection{Controller-Actuator Link}

The controller computes control actions based on $\Tilde{\mathbf{x}}_t$ and sends them to the actuator to be applied to the plant.
Since the \gls{ca} link is wireless, it is susceptible to communication failures caused by poor channel conditions at a given time. 
If the actuator does not receive control action because of that, the system may become unstable as it could drift into unstable regions due to the absence of control. 

Therefore, to ensure system stability, as a precaution method, controller needs to calculate a sequence of  $N_c$ future control actions, denoted as $ [{\mathbf{u}}_t^\tau]_{\tau \in \{t,...,t+N_c \}}$ and transmit them to actuator to cache.
In the event of a \gls{ca} link failure, the actuator applies the cached control action $\hat{\mathbf{u}}_t = {\mathbf{u}}_{t'}^t$ where $t'$ is the most recent time step at which the control actions were successfully received, and the delay satisfies $ t-t' < N_c$.
In that sense, the control action applies to the plant $\Tilde{\mathbf{u}}_t$ at time $t$ can be represented as,
\begin{equation}\label{eq:action_in_actuator}
\Tilde{\mathbf{u}}_t = \begin{cases} 
\mathbf{u}_t, & \text{if \gls{ca} link succeed}, \\
\hat{\mathbf{u}}_{t} & \text{otherwise}.
\end{cases}
\end{equation}

\section{Joint Communication and Control Problem} \label{sec:3}

Maintaining stability in nonlinear and control-non-affine control systems is inherently difficult due to the complexity of their dynamics. 
This difficulty is further exacerbated when \gls{sc} and \gls{ca} links rely on wireless communication, which introduces uncertainties and potential communication failures that compromise control accuracy.
Therefore, achieving stability in such systems demands an efficient approach to ensure reliable control and communication.

%To address the challenges related with absence of states and control actions due to opportunistic \gls{sc} link scheduling and communication link failures, control decisions are typically made based on the $\Tilde{\mathbf{x}}_t$, and the actuator applies the corresponding control action $\Tilde{\mathbf{u}}_t$ on the plant.
%Since $\Tilde{\mathbf{x}_t} = \mathbf{x}_t$ is not always held due to scheduled \gls{sc} transmissions, the controller relies on past data to predict the current state and compute the corresponding control action. 
%Therefore the accuracy of this state prediction directly affects the controller's ability to maintain system stability.

We define a control cost $J_t$ to quantify the effort needed to maintain stability. 
This cost is combined with two costs: a stability cost, which measures the deviation of the system from a desired state, and a control-action cost, which reflects the magnitude of the applied control action.
In traditional control systems, the control cost $J_t$ is typically defined as~\cite{scokaert1998constrained},
\begin{equation}\label{eq:traditional_stabilitycost}
    J_t = (\mathbf{x}_t - \mathbf{x}_{0} )^T \mathbf{Q} (\mathbf{x}_t - \mathbf{x}_{0}) + \mathbf{u}_t^T \mathbf{B} \mathbf{u}_t,
\end{equation}
where $\mathbf{Q} \in \mathbb{S}^D$ and $\mathbf{B} \in \mathbb{R}^{D' \times D'}$ are control system design parameters, and $\mathbf{x}_0$ is the desired state for achieving stability. 
However, in our case, when calculating the control cost $J_t$, we consider the $\Tilde{\mathbf{x}}_t$ instead of the state $\mathbf{x}_t$ because the controller calculates the control action based on it. 
Similarly, since the control effort in $J_t$ is based on the action applied to the plant, we consider the control input $\Tilde{\mathbf{u}}_t$, which is the action applied by the actuator.
Therefore, the control cost $J_t$ for our system is defined as,
\begin{equation}\label{eq:our_stabilitycost}
    J_t = (\Tilde{\mathbf{x}}_t - \mathbf{x}_{0} )^T \mathbf{Q} (\Tilde{\mathbf{x}}_t - \mathbf{x}_{0}) + \Tilde{\mathbf{u}}_t^T \mathbf{B} \Tilde{\mathbf{u}}_t.
\end{equation} 

Due to the scheduled \gls{sc} transmissions, $\Tilde{\mathbf{x}}_t = \mathbf{x}_t$ is not always held and thus the controller relies on past data to predict the current state and compute the corresponding control action.
As a result, accuracy of this state prediction directly affects the controller’s ability to maintain system stability, since $\Tilde{\mathbf{x}}_t$ directly affects both the control action and the resulting cost $J_t$.
Therefore, we must consider about state prediction error, since the accuracy of $J_t$ depends on the accuracy of $\Tilde{\mathbf{x}}_t$.  

The state prediction error is denoted as $E(\Tilde{\mathbf{x}}_t)$ for the system state at the controller, at each time.
Specifically, when $a_t = 1$, the prediction error $E(.) = 0$ because $\Tilde{\mathbf{x}}_t = \mathbf{x}_t$.
Otherwise, the prediction error is influenced by the freshness of system state and the applied control action $\Tilde{\mathbf{u}}_t$ corresponding to the given state $\Tilde{\mathbf{x}}_t$. 

Moreover, \gls{aoi} quantifies how up-to-date the state information is by measuring the time elapsed between successive updates. 
The \gls{aoi} increases over time linearly and can be expressed as:
%Moreover, \gls{aoi} gives a measure of the freshness of state information considering the time interval between consecutive updates. 
%\gls{aoi} increases linearly over time and can be represented as
%
\begin{equation} 
\label{eq:aoi}
\beta_{t} = 1 + (1 - a_t)\beta_{t-1},
\end{equation}
where $\beta_{t}$ corresponds to the \gls{aoi} of the control system at time $t$.
Hence, we can represent the prediction error at time $t$ as,
\begin{equation}\label{eq:prediction_error}
E(\Tilde{\mathbf{x}}_t) = (1 - a_t) E' (\Tilde{\mathbf{u}}_t, \beta_t |\Tilde{\mathbf{x}}_t ),
\end{equation}
where ${E}'(.)$ is a state prediction error function for given $\Tilde{\mathbf{x}}_t$.

Furthermore, calculating $N_c$ consecutive future control actions is particularly challenging for nonlinear, control-non-affine systems due to the inherent complexity of predicting their dynamics. 
Since each control action depends on the system state at its corresponding time step, accurate prediction of future states is essential. 
Therefore, reliable state prediction becomes a critical requirement for computing accurate control actions over the prediction horizon.

Importantly, each \gls{sc} transmission and sensing operation consumes battery power, which is limited. 
In our setting, sensing is assumed to occur continuous and cannot be reduced.
However, the number of \gls{sc} transmissions can be minimized to conserve energy. 
That said, reducing the frequency of \gls{sc} transmissions increases the prediction error, negatively impacting the control system's stability.
Therefore, an optimal \gls{sc} transmission policy must balance the trade-off between prediction accuracy and energy efficiency.

Therefore, the objective of this research is to design communication decision variables such as the \gls{sc} scheduling indicator and the control system decision variable (control action) to minimize \gls{sc} transmissions by ensuring the stability of the control system. In this view, the joint control and communication problem is formulated as,
\begin{subequations}\label{main_problem}
\begin{align}
\minimize_{\mathbf{u}_t , a_{t}} &\,
\frac{1}{T} \left[\sum_{t=1}^{T} J_t + \lambda \sum_{t=1}^{T} a_{t}  \right]\label{eq:mainProblem}\\
\text{s.t.} \quad  &
\label{eq:c}
a_{t} (\gamma_{\text{SC},t} - \gamma_0) \geq 0, \qquad\forall t,\\
\quad  &
%\label{eq:c1}
%\Tilde{\mathbf{u}}_{t} = \{ \mathbf{u}_{t} , \hat{\mathbf{u}}_{t-t'} \}, \qquad\forall t,\\
%\quad  &
\label{eq:c2}
\beta_{t} = 1 + (1-a_{t})\beta_{t-1},  \qquad\forall t, \\
\quad  &
\label{eq:c3}
p_{\text{b},\psi T'} = p_{\text{b},0},  \qquad T < T', \\
\quad  &
\label{eq:c4}
p_{\text{b},t+1} = [p_{\text{b},t} - a_{t} p_{\text{SC},t} - p_{\text{s},t} ]^+, \qquad\forall t, \\
\quad  &
\label{eq:c6}
\Tilde{\mathbf{x}}_t = a_{t} \mathbf{x}_t + (1 - a_{t}) \hat{\mathbf{x}}_t, \qquad\forall t, \\
\quad  &
\label{eq:c7}
\Tilde{\mathbf{u}}_t = a'_{t} \mathbf{u}_t + (1 - a'_{t}) {\mathbf{u}}_{t'}^{t}, \quad\forall t, (t- t') < N_c, \\
\quad  &
\label{eq:c5}
%control and communication constraint
E( \Tilde{\mathbf{x}}_{t} ) \leq \delta,  \qquad\forall t, 
\end{align}
\end{subequations}
%
%$\Tilde{\mathrm{O}}$ is the outage threshold
where $\delta$ is a threshold for prediction error and $\psi \in \mathbb{N}$. 
The $\lambda$ is a weighting parameter which scales the impact of $a_t$, allowing it to be meaningfully added with $J_t$ since the two terms in the objective function may have different magnitudes and unit. 
The status of the \gls{ca} link is represented by the indicator variable $a'$, where $a' = 1$ indicates successful \gls{ca} link at time $t$ and $a'=0$ denotes a failure.

%The $\lambda$ is a weighting parameter which scales the impact of $a_t$ to make it compatible with $J_t$ since the two terms in the objective function may have different magnitudes and units.

The objective in~\eqref{eq:mainProblem} represents the total cost, which is defined as the time-averaged combination of the control cost and number of \gls{sc} transmissions.
The constraint~\eqref{eq:c}  guarantees that the \gls{sc} link must satisfy the \gls{snr} condition $\gamma_{\text{SC},t} \geq \gamma_0$ to enable the successful communication when $a = 1$.
The evolution of \gls{aoi} is defined by constraint~\eqref{eq:c2}, and \gls{aoi} accumulates when the \gls{sc} link is not scheduled.
The constraint~\eqref{eq:c3} ensures that the sensor battery is recharged after $T'$ time, and the constraint~\eqref{eq:c4} represents the battery power update where the battery power is consumed by sensing and \gls{sc} transmission and ensures non-negative battery power.
The system state that controller uses to calculate control action at time $t$ is represented in constraint~\eqref{eq:c6}, 
and the control action that applies on the plant at time $t$ is represented in constraint~\eqref{eq:c7}.
The constraint in~\eqref{eq:c5} limits the expected state prediction error to maintain within a predefined threshold $\delta$, ensuring reliable control decisions.
Furthermore, the evolution of system state $\mathbf{x}_{t}$ defined by the~\eqref{eq:state_dynamic}, is a implicit constraint and it models the nonlinear dynamics of the plant. 
Note that~\eqref{eq:c}-\eqref{eq:c4} constraints are affected directly by scheduling decision variable, and~\eqref{eq:c6}-~\eqref{eq:c5} is affected by both scheduling and control variables. 
While~\eqref{eq:state_dynamic} is only affected by the control decision variable.

\section{Proposed Solution} \label{sec:4}

To address the challenges resulting from the nonlinear, control-non-affine nature of the system dynamics, it is essential to obtain an accurate linear representation of the underlying nonlinear control system.
If such a linear approximation captures the system's dominant behavior with sufficient accuracy, then continuous communication may no longer be necessary to ensure stability.
However, linearization inherently introduces approximation errors induced by the deviation between the nonlinear system dynamics and the linear approximation.
These residual errors can be effectively mitigated by imposing a constraint in~\eqref{eq:c5}.
Once an appropriate linear model is obtained, the problem~\eqref{main_problem} can be decoupled into two subproblems: (i) the control subproblem and (ii) the communication subproblem.

\subsection{Koopman Model for State Prediction and Control Action Planning}

The control subproblem is addressed here, which focuses on minimizing the control cost to ensure stability while satisfying the constraint imposed by the control system's nonlinear dynamics.
The control subproblem is formulated as,
\begin{subequations}  \label{eq:control_sub_problem}
\begin{align}
\underset{  \mathbf{u}_t  }{\minimize}  &\quad
\frac{1}{T}\sum_{t=1}^{T} J_t   \\
\quad 
\text{s.t.}
&\quad 
~\eqref{eq:state_dynamic} , \eqref{eq:c7},
\end{align}
\end{subequations}
with the nonlinear system dynamics constraint.
Here, we do not consider the constraint~\eqref{eq:c6} assuming it is enforced while solving this problem. 

Due to the inherent complexity in predicting states and computing control actions for nonlinear, control-non-affine systems, we utilize the Koopman operator to linearize non-linear system dynamics, which facilitates state prediction and allows for the application of linear control methods such as \gls{lqr}. 
Accordingly, we first review the Koopman operator theory, which forms the mathematical foundation of the proposed solution.
\\

\subsubsection{Preliminary: Koopman Operator with Control}

We consider the nonlinear, control-non-affine control system as~\eqref{eq:state_dynamic}. 
By ignoring system noise, the discrete-time system dynamics are considered as,
\begin{equation}\label{eq:state_dynamic2}
    \mathbf{x}_{t+1} = \mathrm{f}(\mathbf{x}_{t},\mathbf{u}_{t}).
\end{equation}
It can be written follows by de-compositing the dynamics as \cite{hasnain2020steady},
\begin{equation}\label{eq:state_dynamic_decomposition}
    \mathbf{x}_{t+1} = \mathrm{f}_x(\mathbf{x}_{t}) + \mathrm{f}_{xu}(\mathbf{x}_{t}, \mathbf{u}_{t}) + \mathrm{f}_{u}(\mathbf{u}_{t}),
\end{equation}
where $\mathrm{f}_x $ contains terms that only state available, $\mathrm{f}_{xu} $ contains terms that state and control action both available and $\mathrm{f}_u $ contains terms that only control action available.

Then, we consider real valued embedded functions ${g}\colon \mathbb{R}^D \times \mathbb{R}^{D'} \to \mathbb{R}$ which are elements of an infinite-dimensional Hilbert space $\mathcal{H}$~\cite{berberian1999introduction, khosravi2023representer}.
The Koopman operator $\mathcal{K}\colon \mathcal{H} \to \mathcal{H}$ is a linear and infinite-dimensional operator which acts on embedding functions $g$ in Hilbert space as~\cite{kaiser2021data},
\begin{equation}\label{eq:k_operator}
    \mathcal{K}g(\mathbf{x}_t ,\mathbf{u}_t) = g (\mathrm{f} (\mathbf{x}_{t},\mathbf{u}_{t}  ) ,\mathbf{u}_{t+1} ) = g(\mathbf{x}_{t+1} ,\mathbf{u}_{t+1}).
\end{equation} 
According to~\eqref{eq:k_operator}, an infinite-dimensional linear dynamical system is defined by the Koopman operator $\mathcal{K}$, through which $g(\mathbf{x}_t, \mathbf{u}_t)$ is advanced to the next time step $t+1$.

Although $\mathcal{K}$ is linear, there exist computational and representation challenges due to its infinite dimensionality.
Therefore, rather than finding the evolution of all embedding functions belonging to infinite-dimensional $\mathcal{H}$ space, we need to find an invariant subspace spanned by a finite set of embedding functions. 
Since $\mathcal{K}$ is linear, the Koopman invariant subspace can be found via an eigendecomposition. 
The eigenfunctions of the $\mathcal{K}$ yields a set of embedding functions that behave linearly in time.

Towards this, a global linearization expression of a non-linear system dynamics can be obtained by introducing a finite-dimensional Koopman matrix representation $\mathbf{K} \in \mathbb{R}^{q \times q}$ in a given Koopman invariant subspace as,
\begin{equation}\label{eq:k_matrix}
    \mathbf{K}\mathbf{g}(\mathbf{x}_t ,\mathbf{u}_t) = \mathbf{g}(\mathbf{x}_{t+1} ,\mathbf{u}_{t+1} ),
\end{equation}
where $\mathbf{g}$ is the $q$-dimensional vector of which the elements are the Koopman eigenfunctions. 
According to this, if we know concatenated eigenfunctions $\mathbf{g}$, its elementwise inverse $\mathbf{g}^{-1}$ and the Koopman matrix $\mathbf{K}$ for given $\mathbf{x}_t$ and $\mathbf{u}_t$, the future state $\mathbf{x}_{t+1}$ and control action $\mathbf{u}_{t+1}$ can be derived.
But deriving Koopman eigenfunctions from a finite number of embedding functions is challenging. 
So we are using an auto-encoder-based \gls{ml} approach for deriving Koopman eigenfunctions.

Considering the \eqref{eq:state_dynamic_decomposition}, the \eqref{eq:k_matrix} can be rewritten as,
\begin{equation}\label{eq:K_matrix_decomposed}
\begin{bmatrix}
\mathbf{g}_{x}(\mathbf{x}_{t+1}) \\
\mathbf{g}_{xu}(\mathbf{x}_{t+1},\mathbf{u}_{t+1} ) \\
\mathbf{g}_{u}(\mathbf{u}_{t+1})
\end{bmatrix}
=
\begin{bmatrix}
\mathbf{K}_{x} & \mathbf{K}_{xu} & \mathbf{K}_{u}\\
\mathbf{K}_{21} & \mathbf{K}_{22}  & \mathbf{K}_{23} \\
\mathbf{K}_{31} & \mathbf{K}_{32}  & \mathbf{K}_{33}
\end{bmatrix}
\begin{bmatrix}
\mathbf{g}_{x}(\mathbf{x}_t) \\
\mathbf{g}_{xu}(\mathbf{x}_{t},\mathbf{u}_{t} ) \\
\mathbf{g}_{u}(\mathbf{u}_t)
\end{bmatrix}  
\end{equation}
where $\mathbf{g}_x\colon \mathbb{R}^{D}  \to \mathbb{R}^{q} $ contains Koopman eigenfunctions that only related to the system state, $\mathbf{g}_{xu}\colon \mathbb{R}^{D} \oplus \mathbb{R}^{D'}  \to \mathbb{R}^{r}$ contains $r$-dimensional Koopman eigenfunctions that related to the system state and control action mixed terms, and $\mathbf{g}_u\colon \mathbb{R}^{D'}  \to \mathbb{R}^{q'}$ contains $q'$-dimensional Koopman eigenfunctions that only related to the control actions.
In this view, the matrices related to state transition and control action dynamics are represented as $\mathbf{K}_{x} \in \mathbb{R}^{q \times q}$ and $\mathbf{K}_{u} \in \mathbb{R}^{q' \times q'}$.
Furthermore, $ \mathbf{K}_{xu} \in \mathbb{R}^{r \times r}$ is the matrix related to state and control action mixed terms transition. 
The Koopman matrix is represented as $\mathbf{K} \in \mathbb{R}^{(q+r+q') \times (q+r+q')}$ and $\mathbf{K}_{21},..., \mathbf{K}_{33}$ are matrices which represent values for respective embedding functions in the $\mathbf{K}$.

Assuming $\mathbf{u}_t$ as exogenous disturbances without having state space dynamics, first consider the following relation extracted from \eqref{eq:K_matrix_decomposed}:
\begin{equation}\label{eq:embedded_equation}
\mathbf{g}_{x}(\mathbf{x}_{t+1}) = \mathbf{K}_{x}\mathbf{g}_{x}(\mathbf{x}_t)  + \\
\mathbf{K}_{xu}\mathbf{g}_{xu}(\mathbf{x}_t,\mathbf{u}_{t})  + \mathbf{K}_{u} \mathbf{g}_{u}(\mathbf{u}_{t}).
\end{equation}
To avoid the complexity of analysis under the presence of mixed terms of $\mathbf{x}_t$ and $\mathbf{u}_t$ as $\mathbf{g}_{xu}(\mathbf{x}_t,\mathbf{u}_{t})$, we focus on scenarios where $\mathbf{K}_{xu} \approx 0$ \cite{yeung2018koopman}. 
Hence, \eqref{eq:embedded_equation} is rewritten as,
\begin{equation}\label{eq:proposed_main_equation}
\mathbf{g}_{x}(\mathbf{x}_{t+1}) = \mathbf{K}_{x}\mathbf{g}_{x}(\mathbf{x}_t)  + \\
\mathbf{K}_{u} \mathbf{g}_{u}(\mathbf{u}_{t}),
\end{equation}
allowing the derivation of a linear system in the embedding space.
Next, we propose our Koopman-based \gls{dnn} model based on this formulation.
\\

\subsubsection{Proposed Koopman Model Architecture}

\begin{figure}
    \centering
    \includegraphics[width=1\linewidth]{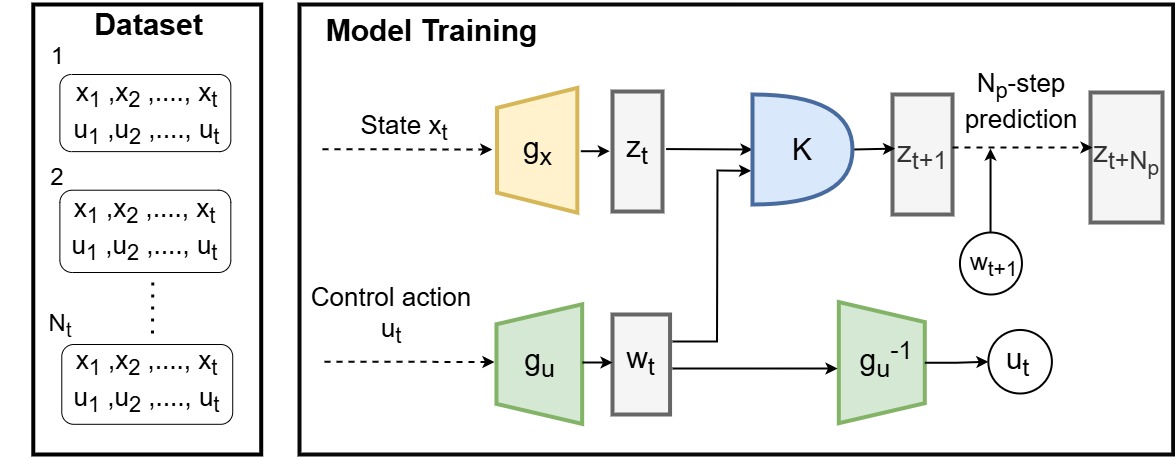}
    \caption{The training process of the proposed Koopman model.}
    \label{fig:model_training}
\end{figure}

\begin{figure}
    \centering
    \includegraphics[width=1\linewidth]{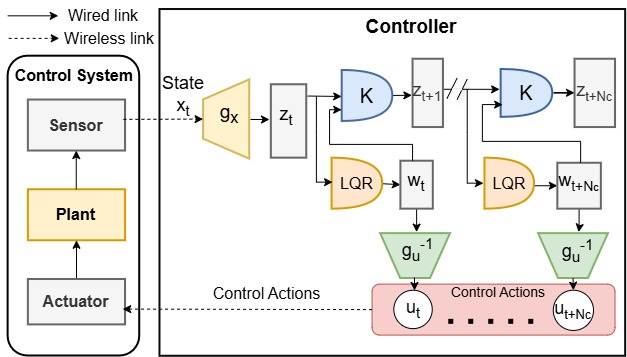}
    \caption{The process of predicting states and calculating control actions using the proposed model.}
    \label{fig:model_inference}
\end{figure}

The proposed \gls{dnn} based Koopman architecture enables the linearization of nonlinear dynamics in the embedded space, allowing for accurate system state predictions and optimal control action computations using linearized dynamics. 
The architecture consists of three main components: (i) encoder layers, which implement the embedding functions $\mathbf{g}_x$ and $\mathbf{g}_u$, (ii) a feedforward layer, responsible for finding finite dimensional Koopman matrix $\mathbf{K}$, and (iii) a decoder layer, which serves as the inverse of $\mathbf{g}_u$. 
The Koopman model operates in two distinct phases: 
(i) \emph{training phase}, which focuses on learning the model parameters using the training data, and 
(ii) \emph{inference phase}, where the trained model is used to predict system states and to compute optimal control actions.

The training phase focuses on learning the Koopman-based \gls{dnn} model that enables the linearization of nonlinear system dynamics in the embedding (latent) space. 
The proposed model architecture consists of five key components:

\begin{enumerate}
     \item State embedding layer $\mathbf{g}_x(.)$: This layer maps the original system state $\mathbf{x}_t$ to an embedding state representation $\mathbf{z}_t$. %, which serves as the embedding state. 
     It incorporates a \gls{nn} encoder $g_\phi\colon \mathbb{R}^D \to \mathbb{R}^{q-D}$ which is parameterized by $\phi$ and responsible for encoding $\mathbf{x}_t$ to an encoded state ${g_\phi(\mathbf{x}_t)}$.
     The final embedding state $\mathbf{z}_t$ is obtained by concatenating the original system state $\mathbf{x}_t$ with its encoded representation, formulated as:
        \begin{equation}\label{eq:state_encode}
            \mathbf{z}_t = \mathbf{g}_x(\mathbf{x}_t) = \begin{bmatrix} \mathbf{x}_t, {g_\phi}(\mathbf{x}_t) \end{bmatrix}^T.
        \end{equation}

    Since $\mathbf{x}_t$ is directly included in $\mathbf{z}_t$, the model cannot ignore the original state information. This structure ensures that $\mathbf{g}_x(.)$ learns a meaningful mapping from the original state space to the embedding space while preserving system dynamics.

    \item Control action embedding layer $\mathbf{g}_u(.)$: This layer maps the original control action $\mathbf{u}_t$ to the embedding space control action $\mathbf{w}_t$, ensuring that the control inputs align with the transformed embedding space. 
    We use \gls{nn} decoder $g_\mu\colon \mathbb{R}^{D'} \to \mathbb{R}^{q'}$ parameterized by $\mu$ such that $\mathbf{g}_u(\mathbf{u}_t) = {g}_{\mu}(\mathbf{u}_t)$.
     
    \item Koopman operator layer $\mathbf{K}$: The Koopman operator layer captures the linear state dynamics in the embedding space, allowing embedding state predictions.

    \item Control action decoding layer $\mathbf{g}^{-1}_u(.)$: This layer maps the latent space control actions  $\mathbf{w}_t$ back to the original control action space. It serves as the inverse function of $\mathbf{g}_u$, ensuring that control actions can be accurately reconstructed. Furthermore, $\rho$ is used to parameterize $\mathbf{g}^{-1}_u$ where ${g}_\rho \colon \mathbb{R}^{q'} \to \mathbb{R}^{D'}$ is a \gls{nn} decoder such that $\mathbf{g}_u^{-1}(\mathbf{u}_t) = {g}_{\rho}(\mathbf{u}_t)$. This is important, as the optimal control action is derived in the embedding space, and this mapping allows it to be projected in the original space.   

\end{enumerate}

Since the Koopman operator captures the linearized system dynamics in the embedding space, it enables future state predictions when the current system state and the corresponding control actions are known. 
This allows the model to iteratively predict system states over future consecutive time steps. 

To train the proposed Koopman-based model, we utilize a dataset generated by applying random control actions to given system states and recording the resulting next states. 
This dataset consists of recorded samples from multiple trajectories and $N_t$ number of steps for each trajectory.
A particular sample contains the original system states $\mathbf{x}_t$, applied control action $\mathbf{u}_t$, and the corresponding next states $\mathbf{x}_{t+1}$.
Using this data, we train the model parameters of the ${g}_\phi(.)$, ${g}_\mu(.)$, $\mathbf{K}$, and ${g}_\rho(.)$ layers end-to-end.

The training process is guided by a
loss function that accounts for state evolution and control action reconstruction to ensure accurate long-term $N_p$ time steps predictions.
Given the dataset with sets of system states $\mathcal{X}_{k} \in \mathbb{R}^{N_t \times D}$ and control actions $\mathcal{U}_{k} \in \mathbb{R}^{N_t \times D'}$ where $ k = 0,1,...,N_p$, the model computes the sets of embedding space system states $\mathcal{Z}_{k} \in \mathbb{R}^{N_t \times q}$ and predicted embedding space system states set over $N_p$ time steps as $\hat{\mathcal{Z}}_{k} \in \mathbb{R}^{N_t \times q}$ where $k = 1,...,N_p$.
For the control actions also computes the embedding space control action set  $\mathcal{W}_{k} \in \mathbb{R}^{N_t \times q'}$ and decoded control action set $\hat{\mathcal{U}}_{k} \in \mathbb{R}^{N_t \times D'}$.
The loss function $L_{\text{MSE}}$ is formulated using \gls{mse} with two losses as: 
\begin{equation}\label{eq:loss_function}
    L_{\text{MSE}}(\phi, \mu, \rho) = \sum^{N_p}_{k} {M_e}(\mathcal{Z}_{k}, \hat{\mathcal{Z}}_k) + {M_e}(\mathcal{U}_k, \hat{\mathcal{U}}_k),
\end{equation}
where $M_e(\mathcal{Z}_{k}, \hat{\mathcal{Z}}_k)$ is the embedding state prediction loss~\cite{ma2022principles}, that captures the error between the predicted and actual embedding states over multiple time steps, and $M_e(\mathcal{U}_k, \hat{\mathcal{U}}_k)$ is the control action reconstruction loss, which ensures proper inverse mapping of control inputs from the embedding space.
By jointly optimizing these losses, the model effectively learns a structured embedding representation where system dynamics evolve linearly.

The loss function, which is computed in the embedding space, still has a direct impact on the accuracy of the state embedding because since $\mathbf{x}_t$  is explicitly part of $\mathbf{z}_t$ and ensures the good training of the state embedding layer $\mathbf{g}_x(.)$. 
Any error in predicting $\mathbf{z}_{t+1}$ inherently affects the prediction of $\mathbf{x}_{t+1}$, ensuring that the learnable part of $\mathbf{g}_x(\mathbf{x})$ is properly optimized by~\eqref{eq:loss_function}.

The trained Koopman model is then utilized in the inference phase to achieve the desired control objectives. 
During inference, trained ${g}_\phi(.)$, $\mathbf{K}$, and ${g}_\rho(.)$ layers are used along with linear control method \gls{lqr} to predict states and plan control actions.

To apply the linear control method to find the optimal control action to stabilize the system, the aforementioned control subproblem \eqref{eq:control_sub_problem} is reformulated as,
\begin{subequations}  \label{eq:control_sub_problem_latent}
\begin{align}
\underset{  \mathbf{w}_t  }{\minimize} \label{subeq:lqr_cost1}
\frac{1}{T}\sum_{t=1}^{T} ( ({\mathbf{z}_t - \mathbf{z}_{0} } )^T \hat{\mathbf{Q}} (\mathbf{z}_t - \mathbf{z}_{0}) + \mathbf{{w}}_t^T \hat{\mathbf{B}} \mathbf{{w}}_t)   \\
\quad 
\text{s.t.} 
\quad \label{subeq:lqr_cost2}
\mathbf{z}_{t+1} = \mathbf{K}_{x}\mathbf{z}_t  + \mathbf{K}_{u} \mathbf{w}_{t}, \qquad\qquad\quad \forall t, 
\end{align}
\end{subequations}
where $\hat{\mathbf{Q}}  \in \mathbb{S}^q$ and $\hat{\mathbf{B}}  \in \mathbb{R}$ are embedding space control system design parameters which are designed such that $({\mathbf{z}_t - \mathbf{z}_{0} } )^T \hat{\mathbf{Q}} (\mathbf{z}_t - \mathbf{z}_{0}) \approx ({\mathbf{x}_t - \mathbf{x}_{0} } )^T {\mathbf{Q}} (\mathbf{x}_t - \mathbf{x}_{0})$ and $\mathbf{{w}}_t^T \hat{\mathbf{B}} \mathbf{{w}}_t) \approx \mathbf{{u}}_t^T {\mathbf{B}} \mathbf{{u}}_t)$.
The desired embedding space system state is represented as $\mathbf{z}_{0} = \mathbf{g}_x(\mathbf{x}_{0})$.
The constraint~\eqref{subeq:lqr_cost2} represents the dynamics in the embedding space.
Since the system dynamics in the embedding space are linear, the constraint~\eqref{subeq:lqr_cost2} defines a linear relationship between $\mathbf{z}_{t}$ and $\mathbf{w}_{t}$.

Since this problem~\eqref{eq:control_sub_problem_latent} is with linear quadratic cost function and linear constraint, optimal embedding space control action can be derived using \gls{lqr}.
Therefore, optimal embedding space control action can be derived as,
\begin{equation}\label{eq:lqr}
    \mathbf{w}^*_t = - K_{\text{LQR}}(\mathbf{z}_t - \mathbf{z}_{0}),
\end{equation}
where $K_{\text{LQR}}$ is the feedback gain matrix that is derived from the solution of the Riccati equation~\cite{lancaster1995algebraic}.
Thereafter the original space control action can be found using ${g}_\rho(.)$ as,
\begin{equation}\label{eq:optimal_u}
    \mathbf{u}^*_t = {g}_\rho(\mathbf{w}_t^*).
\end{equation}

The proposed model can be applied under two distinct scenarios.
When the sensor is scheduled, the model processes received sensor data by mapping it into the embedding space using the $\mathbf{g}_x(.)$, predicting the next state using the Koopman layer $\mathbf{K}$, calculating the optimal embedding space control action using \gls{lqr} and, mapping embedding space control action to original space optimal control action using $\mathbf{g}^{-1}_u(.)$.
When the sensor is not scheduled, the model computes the control actions based on the past system states by iteratively predicting future embedding space states using the Koopman layer $\mathbf{K}$.

Furthermore, the predicted system states can be used to plan a sequence of future control actions over a prediction horizon. 
Therefore, in our proposed solution, the controller computes optimal control inputs for a consecutive $N_c$ number of future time steps, sends them to the actuator each time, and caches them in the actuator. 
This mechanism provides resilience against \gls{ca} link communication failures, as the actuator can autonomously execute the cached control actions when connectivity is compromised.

\subsection{Scheduling algorithm based on Lyapunov Optimization}

In this section, we address the communication subproblem, which aims to minimize the number of \gls{sc} transmissions while ensuring stability of system.
Lyapunov optimization is utilized since this is a time-averaged dynamic stochastic optimization problem.
Thereafter, the communication subproblem is formulated as,
\begin{subequations}  \label{eq:comm_sub_problem}
\begin{align}
\underset{ a_{t} }{\minimize}  &\quad
\frac{\lambda}{T}\sum_{t=1}^{T} a_{t}   \label{subeq:comm_sub_1}\\
\quad 
\text{s.t.}
\label{subeq:comm_sub_2}
&\quad 
\eqref{eq:c}-\eqref{eq:c5},
\end{align}
\end{subequations}
including constraints related to communication reliability and prediction error constraints.

Since an optimal control action $\mathbf{u}_t^*$ is capable of being determined using Koopman model, $E(.)$ is possible to be interpreted as a function of the last received system state from the sensor $\mathbf{x}_{t-{\beta_t}}$ and the \gls{aoi} $\beta_t$.
Therefore, given $\mathbf{u}^*_t$ is known, we reformulate the constraint~\eqref{eq:c5} by expressing the expected state prediction error as a function of $\mathbf{x}_{t-{\beta_t}}$ and $\beta_t$ under the optimal control action.
Hence, $E(.)$ can be reformulated as,
\begin{equation}\label{eq:prediction_error}
    E(\Tilde{\mathbf{x}}_t) = (1 - a_t) \mathcal{E}(\mathbf{x}_{t-{\beta_t}}, \beta_t | \mathbf{u}^*_t ), 
\end{equation}
where $\mathcal{E}(.)$ is a state prediction error function for given $\mathbf{u}^*_t$. 

%It is because finding a closed-form expression for $E(\mathbf{u}_t,\beta_t | \mathbf{x}_t)$ is difficult when considering all the possibilities of $\mathbf{u}_t$.

Thereafter, we find the closed-form approximation for the prediction error function $\mathcal{E}(.)$ by using a least squares polynomial approximation, which is a data-driven approach.
The polynomial degree is selected based on the one that yields the minimum difference between the actual prediction error $E(.)$ and the approximated prediction error $\mathcal{E}(.)$.
Here, assuming a second-degree polynomial approximation with coefficients  $\boldsymbol{\alpha} = [ \alpha_i ]_{i=\{1,...,5\}}$, the approximated prediction error can be represented as,
\begin{equation}\label{eq:polynomial_approx_degree2}
   \mathcal{E}(\mathbf{x}_t,\beta_t) = { \boldsymbol{\alpha} }^T [ \| \mathbf{x}_{t} \|, \beta_{t}, \| \mathbf{x}_{t} \|^2, \beta_{t}^2, \| \mathbf{x}_{t} \|\beta_{t} ].   
   %\alpha_1 \| \mathbf{x}_{t} \| +  \alpha_2\beta_{t} + \alpha_3{\| \mathbf{x}_{t} \| }_{t}^2 + \alpha_4\beta_{t}^2 + \alpha_5{\| \mathbf{x}_{t} \| }_{t}\beta_{t}
\end{equation}

In order to capture the evolution of a number of \gls{sc} transmissions over time by using virtual queues, the problem objective is reformulated as minimizing the expected number of \gls{sc} transmissions at each time step.
Following that, the reformulated problem is represented as,
\begin{subequations} \label{eq:exp_trans_2}
\begin{align}
\underset{ a_{t}, \Gamma_{t}\l }{\minimize} &\quad
\mathbb{E}[\Gamma_{t}] \label{subeq:exp_trans2_1} \\
\quad 
\text{s.t.}
\label{subeq:expected_transmission2}
&\quad
a_{t} \in [0,1],  \qquad \forall t, \\
& \quad \label{subeq:expected_transmission3}
\mathcal{E}(\mathbf{x}_{t-\beta_t},\beta_t | \mathbf{u}_t^*) \leq \delta, \qquad \forall t, \\
&\quad \label{subeq:exp_trans2_2}
a_{t} \leq \Gamma_t, \quad\qquad \forall t, \\
&\quad \label{subeq:exp_trans2_3}
0 \leq \Gamma_t \leq 1, \quad\qquad \forall t, \\
& \quad
\eqref{eq:c}-\eqref{eq:c7},
\end{align}
\end{subequations}
where $\Gamma$ is the auxiliary variable for defining an upper bound for $a_t$. 
The constraint~\eqref{subeq:expected_transmission2} is used to relax the scheduling decision by bounding between 0 and 1, and the constraint~\eqref{subeq:exp_trans2_3} is introduced to define the upper and lower bounds of the auxilary variable considering bounds of $a_t$.
The problem~\eqref{eq:exp_trans_2} is equivalent to the problem~\eqref{eq:comm_sub_problem} and an optimal solution can be derived to the~\eqref{eq:comm_sub_problem} by solving problem~\eqref{eq:exp_trans_2}. 

A virtual queue is introduced to represent the number of transmissions at each time, and  its corresponding dynamics are represented by,
\begin{equation}\label{eq:virtual_queue}
    Q_{a,t+1} = \max\{Q_{a,t} - \Gamma_t , 0\} + a_{t}, 
\end{equation}
where $Q_{a,0} = 0$ and $a_{t}$ is optimized at each time $t$. 
We consider $\Gamma_t$ as the upper bound for scheduling decisions and $a_t$ as the arrival rate for this virtual queue.
By ensuring the stability of the virtual queue, the constraint~\eqref{subeq:exp_trans2_2} can be satisfied because the lower bound of constraint~\eqref{eq:virtual_queue} is the arrival rate of the virtual queue, while the upper bound of constraint~\eqref{eq:virtual_queue} is the service rate of the virtual queue.

Thereafter Lyapunov optimization is used to solve the problem~\eqref{eq:exp_trans_2}. 
The Lyapunov function captures the system's current state regarding battery power $p_{b,t}$, \gls{aoi} $\beta_t$, and scheduling decisions $Q_{a,t}$. 
Minimizing the Lyapunov drift-plus-penalty ensures that the system remains stable while optimizing transmission decisions.
Therefore, the Lyapunov function is defined as,
\begin{equation}\label{eq:lyapunov_function}
    L(Q_t) = \frac{1}{2} (p_{b,t}^2 + \beta_{t-1}^2 + Q_{a,t}^2 ),
\end{equation}
and then the Lyapunov drift is formulated as,
\begin{equation}\label{eq:lyapunov_drift}
\begin{split}
    \Delta L=&  L(Q_{t+1}) - L(Q_t) \\
%    =& \frac{1}{2} [(p_{b,t+1}^2 + \beta_{t}^2 + Q_{a,t+1}^2) - (p_{b,t}^2 + \beta_{t-1}^2 + Q_{a,t}^2) ]\\
    =& \frac{1}{2}[1 - 2p_{b,t}(a_{t} p_{\text{SC},t} + p_{s,t} ) +2\beta_{t-1}(1-a_t)  \\  & +2Q_{a,t}(a_{t} -\Gamma_t) +(a_{t} p_{\text{SC},t}
     + p_{s,t})^2 -a_{t}\beta^2_{t-1}(2-a_t)\\
    & + (a_{t} -\Gamma_t)^2 (p_{b,t}^2 + \beta_{t-1}^2 + Q_{a,t+1}^2) ].
\end{split}
\end{equation}
Afterward, drift-plus-penalty is formulated as $ \mathbb{E} [\Delta L] + V \mathbb{E}[\Gamma_t]$ where $V \geq 0$ is nonnegative weight that affects a performance tradeoff.
$\mathbb{E}[.]$ provides the expected value of a given variable. 
We can consider square terms constants and formulate the above drift-plus-penalty equation as the following inequality.
\begin{equation} \label{eq:lyapunov_drift_plus_penalty}
\begin{split}
\mathbb{E}[\Delta L] + V \mathbb{E}[\Gamma_t] & \leq  B + \mathbb{E}[V \Gamma_t + Q_{a,t}(a_{t} -\Gamma_t) \\
& - \quad p_{b,t}(a_{t} p_{\text{SC},t} + p_{s,t} ) +\beta_{t-1}(1-a_t)].   
\end{split}
\end{equation}
Henceforth, we omit the constant $B$ in~\eqref{eq:lyapunov_drift_plus_penalty} since it has no impact on the system performance in the Lyapunov optimization.
Therefore, an optimal solution to the problem~\eqref{eq:exp_trans_2} is derived by minimizing the upper-bound of~\eqref{eq:lyapunov_drift_plus_penalty} at each time $t$ as,
\begin{subequations}\label{eq:final_problem}
\begin{align}
\underset{ a_{t}, \Gamma_{t}\l }{\minimize} &\quad
V \Gamma_t + Q_{a,t}(a_{t} -\Gamma_t) \\
& - \quad p_{b,t}(a_{t} p_{\text{SC},t} + p_{s,t} ) +\beta_{t-1}(1-a_t) \\
\quad 
\text{s.t.}
&\quad 
\eqref{subeq:expected_transmission2}-\eqref{subeq:exp_trans2_3},\eqref{eq:c}-\eqref{eq:c7}.
\end{align}
\end{subequations}

Note that except ~\eqref{subeq:expected_transmission3}, the objective and the constraints are linear. 
By substituting~\eqref{eq:aoi} into ~\eqref{subeq:expected_transmission3}, the expression can be written in the form $c_1 a^2 + c_2 a + c_3  \leq 0$, where $c_1$ and $c_2$ are the coefficients of $a^2$ and $a$, respectively, and $c_3$ is a constant term. 
The values of $c_1$, $c_2$ and $c_3$ are obtained by simplifying the corresponding coefficient terms in \eqref{eq:polynomial_approx_degree2}.
If $c_1 \geq 0$, the constraint~\eqref{subeq:expected_transmission3} will be a convex function and can be solved as a convex problem. 
Otherwise, we can transform~\eqref{subeq:expected_transmission3} as a linear matrix inequality as,
\begin{equation}
    \begin{bmatrix}
        a_{t} \\
        1
    \end{bmatrix}^T 
    \begin{bmatrix}
        c_1           &   \frac{c_2}{2} \\
        \frac{c_2}{2} &    c_3
    \end{bmatrix}
    \begin{bmatrix}
        a_{t} \\
        1
    \end{bmatrix} \geq 0
\end{equation}

Therefore, for function~\eqref{subeq:expected_transmission3} to be convex, the following matrix is relaxed as 
$\begin{bmatrix}
    c_1           &   \frac{c_2}{2} \\
    \frac{c_2}{2} &    c_3
\end{bmatrix} 
\succcurlyeq 0 $ and then it leads~\eqref{subeq:expected_transmission3} constraint to an \gls{sdp} constraint which would help to solve this problem as a convex problem.

\section{Results}\label{sec:5}

\begin{table}[!t]
    \centering
    \caption{Simulation parameter values}
    \label{tab:simulation_par}
    \begin{tabular}{|c|c|c|c|}
        \hline
        Parameter& Value & Parameter& Value \\ [0.5ex] 
        \hline
         $m_1$ & $2$~kg  & $V $ & $ 10$ \\ \hline        
         $m_2$ & $2$~kg & $\lambda $ & $ 1$ \\ \hline      
         $j_1$ & $0.5$~kg & $\mathbf{x}_0$ &$[0,0,0,0]$ \\ \hline
         $j_2$ & $0.5$~kg & $N_0 $ & $ -168~\text{dBmHz}^{-1}$ \\ \hline
         $g' $ & $ 10~\text{ms}^{-2}$ & $N_c $ & $ 10$ \\ \hline
         $l'$ & $ 0.5$~m  & $\omega $ & $ 2.4$~GHz \\ \hline
         $b$ & $ 0.4$~m  &  $\kappa  $ & $ 10$ \\ \hline
         $k' $ & $2~\text{Nm}^{-1}$ & $\delta $ & $ 0.3$ \\ \hline         
         $s $ & $ 0.5$~m & $\Tilde{O} $ & $ 0.001$ \\ \hline
         $\mathbf{Q} $ & $ \operatorname{diag} (20, 0.01, 5, 0.01)$  & $p_{s,t}$  & $ 10^{-5}~\text{W} $ \\ \hline
         $ \mathbf{B}  $ & $ 0.001\mathbf{I}$  & $p_{b,0}$  & $ 1~\text{W} $ \\
                 
         \hline
    \end{tabular}
\end{table}

To validate our theoretical approach, in this section we provide numerical results using two experimental setups: the control-affine cartpole system~\cite{brockman2016openai} and the nonlinear control-non-affine double pendulum setting, which is connected by a spring~\cite{karimi2009decentralized}.
The cartpole dynamics are the same as in the OpenAI gym environment~\cite{brockman2016openai}. 
In the double pendulum system, the state is described by $\mathbf{x} = [\theta_1, \dot\theta_1, \theta_2, \dot\theta_2]$ where $\theta_i$ and $\dot\theta_i$ are angle from vertical axis and angular velocity of pendulum $i \in \{1,2\}$. 
The torque applied to each of the pendulums $\mathbf{u} = [u_1, u_2]$ is the control action.
Hence, the system dynamics are represented as,
\begin{align}
    \frac{d \theta_i}{dt} = &\dot\theta_i,  \qquad i \in \{1,2\}, \\
    \frac{d \dot\theta_1}{dt} =& \left( \frac{m_1 g' s}{j_1} - \frac{k' s^2}{4 j_1}\right) \sin({\theta_1}) + \frac{k' s}{2 j_1}(l' - b)\\ \nonumber &+ \frac{h'(u_1)}{j_1} + \frac{k' s^2}{4 j_1}\sin(\theta_2),\\
    \frac{d \dot\theta_2}{dt} =& \left( \frac{m_2 g' s}{j_2} + \frac{k' s^2}{4 j_2}\right) \sin({\theta_2}) - \frac{k' s}{2 j_2}(l' - b) \\ \nonumber &+ \frac{h'(u_2)}{j_2} + \frac{k' s^2}{4 j_2}\sin(\theta_1),
\end{align} 
where $m_1$ and $m_2$ are masses of each pendulum, $j_1 $ and $j_2$ are the moments of inertia, $l'$ is the spring length, $b $ is distance between two pendulums, $k'$ is spring constant, $g' $ is gravity and $s $ is the pendulum height.
Two different nonlinear control action modelings are considered by $h'({u}_i) = \tanh({u}_i)$ and $h'({u}_i) = {u}_i - \frac{1}{3}{u}_i^3$ where $i \in \{1,2\}$. 
Furthermore, the Table \ref{tab:simulation_par} shows the default parameter values used in simulations.

%For each simulation, 1000 time steps have been carried out.
%Furthermore, all analyses are based on averaging data collected over 100 simulations.

\subsection{Model training}

The control system simulations were developed using Python to facilitate the generation of training data for the Koopman operator-based model. 
In data generation, random control actions were applied to various system states, and the corresponding next states were computed. 
To construct the dataset, this process was repeated across over $5000$ trajectories, each with $2000$ timesteps.

Deep Koopman model is implemented in Python using the PyTorch framework~\cite{imambi2021pytorch}.
The model was trained to minimize the loss function defined in~\eqref{eq:loss_function} using the generated dataset. 
We use the Adam optimizer~\cite{jais2019adam} with a batch size of $10^{3}$ and a learning rate of $10^{-3}$.
The model architecture consists of four \gls{nn} modules as represents in Table~\ref{tab:Model_simu_para}, each implemented as a fully connected feedforward network.

\begin{table}[!t]
 \caption{\gls{nn}-related parameters in the format of number of neurons in input, hidden, and output layers followed by the activation function.}
\centering
\label{tab:Model_simu_para}
\begin{tabular}{ |c|c| }
\hline
Layer & Parameter values \\[0.5ex]
\hline
{State embedding layer} & (4, 128, 128, 128, 20, ReLU) \\
\hline
{Control action embedding layer} & (2, 128, 128, 128, 2, ReLU) \\ 
\hline
{Koopman operator layer 1} & (24,24) \\ 
\hline
{Koopman operator layer 2} & (2,24) \\ 
\hline
{Control action decoding layer} & (2, 128, 128, 128, 2, ReLU) \\ 
\hline
\end{tabular}
\end{table}

To model the prediction error function ($\mathcal{E}$), we collected $10^{5}$ prediction error values from the Koopman model for given values of $\mathbf{x}_t$ and $\beta$, with $\beta$ limited to a maximum of $30$, and the initial $\mathbf{x}_t$ randomly selected. 
Thereafter, a linear regression model is subsequently trained on these samples to estimate prediction error coefficients $\boldsymbol{\alpha}$. 
%The corresponding coefficient values $\boldsymbol{\alpha}$ are calculated for the Koopman model.

The communication channel was modeled as a Rician fading channel with a Rician factor $\kappa = 10$. 
Transmission power $p_i$ is allocated to achieve a target outage probability $\Tilde{O} = 10^{-3}$ and $\gamma_0 = 20$\,dB.
The performance of the proposed model is further evaluated under varying outage probability thresholds from $10^{-1}$ to $10^{-4}$.

%explain baselines
As for comparing our Koopman model's performance with the existing literature, we consider Koopman models in~\cite{shi2022deep}, \gls{dkuc} model, which considers embedded space control action as the same as the original space control action, and \gls{dkac} model, which considers control systems as control-affine systems. 
Furthermore, To evaluate the benefits of planning control actions over multiple time steps, we compare the proposed solution with two baseline scenarios: B1 and B2.
During \gls{ca} link failures no control is applied with B1 while the most recently received control action is reused with B2.
For a fair comparison, we have used our scheduling algorithm in sensor scheduling in the baselines.

\subsection{Performance evaluation}

\subsubsection{Prediction Accuracy }

\begin{figure}
    \centering
    \includegraphics[width=1\linewidth]{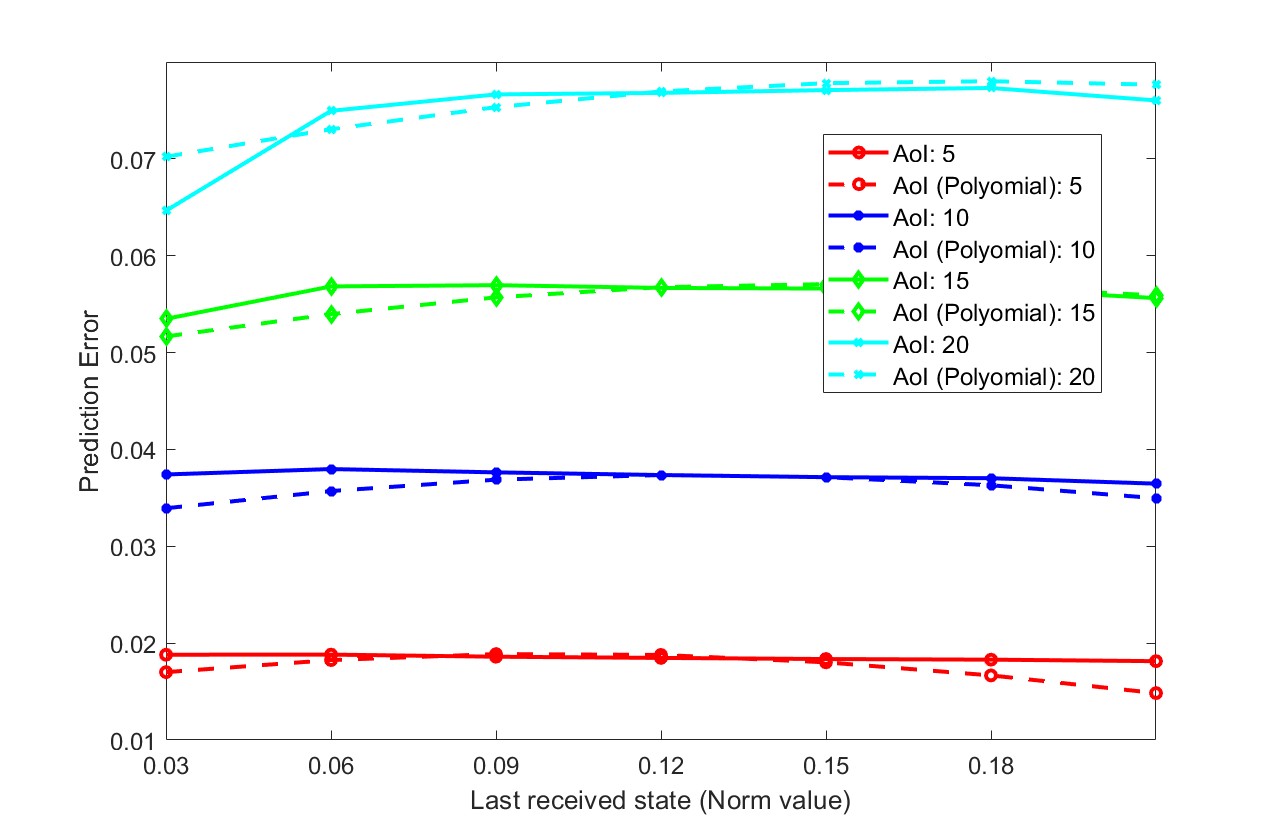}
    \caption{Comparison between polynomial approximated error and actual prediction error using a degree 2 approximation.}
    \label{fig:poly_error_vs_real_error}
\end{figure}

Evaluating the state prediction accuracy of the proposed system is essential, especially in scenarios where successive \gls{sc} transmissions are missing, leading to an increased \gls{aoi}.
In our approach, the state prediction error $E(.)$ is approximated using \eqref{eq:prediction_error} at the sensor.
To determine the appropriate degree for the polynomial approximation, we first analyze the difference between the actual prediction error and its polynomial approximation, expressed as $E(.) - \mathcal{E}(.) $, across various polynomial degrees ($1, 2$, and $3$).
Here, double pendulum is considered as the control system and analysis is performed considering different \gls{aoi} values ranging from $1$ to $30$ and randomly selected last received state $\mathbf{x}_t$.
According to the results, a degree $2$ polynomial provides the most accurate approximation, and therefore it is selected as the polynomial degree.

For further evaluation, we analyze the actual prediction error and approximated prediction error separately for degree 2 polynomial. 
Therefore, Fig.~\ref {fig:poly_error_vs_real_error} illustrates the prediction error, which is calculated using the polynomial approximation $\mathcal{E}(.)$ and the real prediction error $E(.)$ for different \gls{aoi} $(5, 10, 15,$ and $20)$ values.
The results indicate that the prediction error generally increases with higher \gls{aoi}. 
This behavior is expected because a higher \gls{aoi} implies that the model relies on increasingly outdated information, making accurate state estimation more difficult.
However, the difference between the actual and approximated prediction error is significantly low and therefore it indicates that the linear approximation is capable of accurately model the behavior of prediction error across the full state range. 

\subsubsection{Stability comparison}

\begin{figure}
    \centering
    \includegraphics[width=1.1\linewidth]{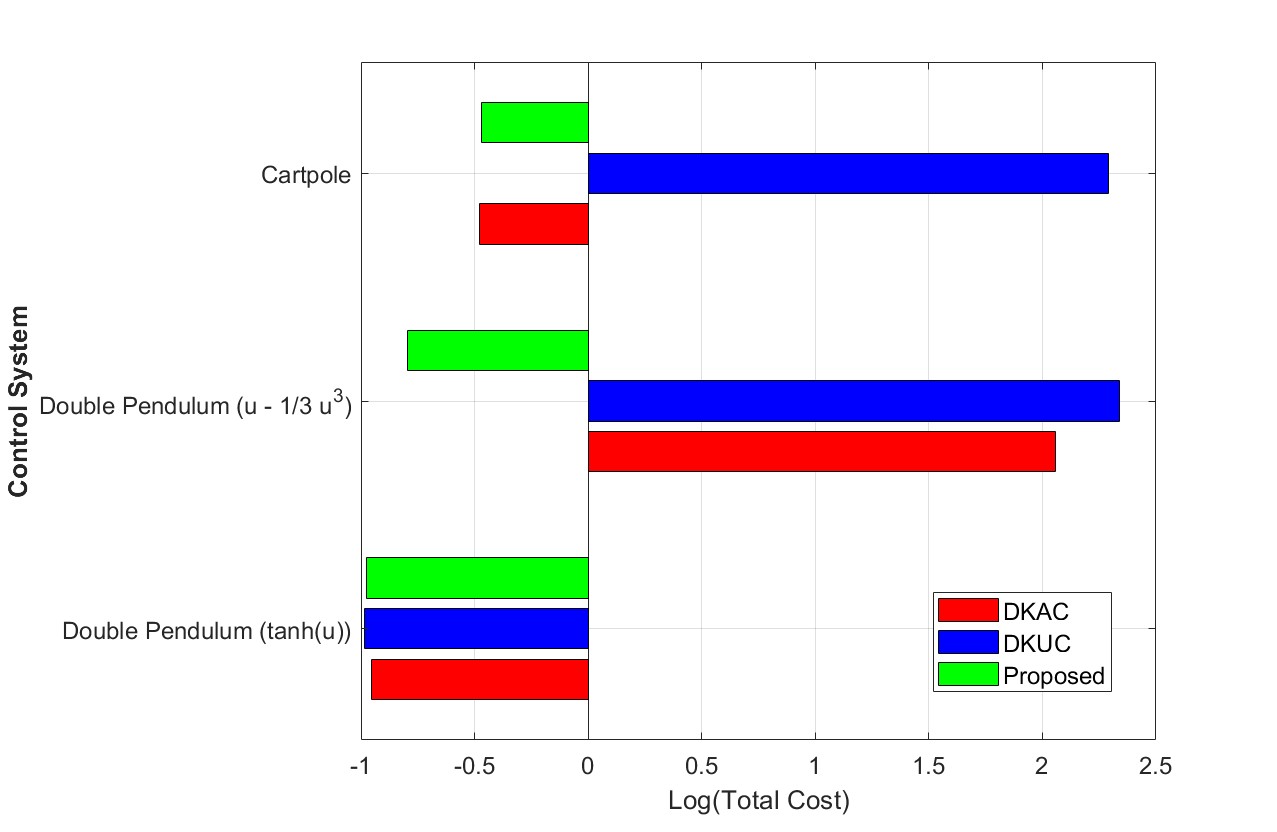}
    \caption{Log value of total cost in~\eqref{eq:mainProblem} across different control systems, comparing the proposed model with the baselines.}
    \label{fig:total_cost}
\end{figure}

Fig.~\ref{fig:total_cost} illustrates the logarithmic values of the total cost in~\eqref{eq:mainProblem} for each control system, comparing the proposed method with two baseline models,  \gls{dkac} and \gls{dkuc}. 
The results are averaged over $100$ simulations, each running for $1000$ timesteps, with $V = 10$ and $h'(.) = \tanh(.)$.

As observed from the results, in the double pendulum scenario where the nonlinear form of control input is applied using $\tanh(.)$ function, all three methods, achieve stability with relatively low costs $0.1116, 0.1040$ and $0.1052$ respectively.
This suggests that baseline models can effectively approximate control actions when the input is constrained within the bounded range of $(-1, 1)$, as imposed by the $\tanh(.)$ function. 
In contrast, for the cartpole system, only the \gls{dkac} and the proposed method achieve low costs of $0.3328$ and $0.34$, respectively, while the \gls{dkuc} model incurs a significantly higher cost of $194.8$.
This difference highlights \gls{dkuc}’s inability to handle the control-affine dynamics of the cartpole system effectively.
Notably, in the more challenging double pendulum setting where the control input follows the form ${u} - \frac{1}{3}{u}^3$, only the proposed method maintains a low total cost $0.1607$.
This corresponds to a proposed model's total cost reduction is $99.9\%$ compared to both baseline models.
Specially, the time-averaged control costs for \gls{dkac}, \gls{dkuc}, and the proposed method are $114.04$, $217.93$ and $0.057$, respectively, while the time-averaged number of \gls{sc} transmissions remains at $0.104$ across all three methods.
This indicates its capability in stabilizing systems with complex nonlinear and control-non-affine dynamics.
Overall, these results demonstrate that the proposed model is robust across a wide range of control scenarios. 
It consistently stabilizes control-affine and control-non-affine nonlinear systems, outperforming baseline methods.

%\subsubsection{Outage probability vs No of \gls{sc} transmissions}

%\begin{figure}
    %\centering
    %\includegraphics[width=1\linewidth]{Figures/outage_vs_a.jpg}
    %\caption{Number of transmissions vs Outage probability}
    %\label{fig:outage_vs_a}
%\end{figure}

%Fig.~\ref{fig:outage_vs_a} illustrates the relationship between the outage probability threshold and the average number of \gls{sc} transmissions, considering  \gls{snr} threshold as $20~dBm$. 
%As depicted, the number of SC transmissions remains relatively low and stable for strict outage thresholds $( 10^{-4} $ to $ 10^{-2})$. However, as the threshold increases further, particularly at $10^{-1}$, a sharp rise in the number of transmissions is observed, reaching over $450$ on average.
%This trend can be attributed to the system’s effort to maintain stability under relaxed outage threshold. 
%When the outage probability threshold is high, the system allows a greater tolerance for unreliable transmissions. 
%Consequently, to counteract the increased likelihood of outages and ensure robust system performance, the \gls{sc} link scheduler initiates more \gls{sc} transmissions.
%This trend highlights the trade-off between transmission reliability and transmission frequency in our proposed approach.
%Low outage thresholds results in fewer but more reliable \gls{sc} transmissions, conserving battery power.
%High outage thresholds demands frequent transmissions to counterbalance the reduced individual transmission reliability which causes for high battery power.

\subsubsection{Planning control actions when consecutive \gls{ca} link failures}

\begin{figure}
    \centering
    \includegraphics[width=1\linewidth]{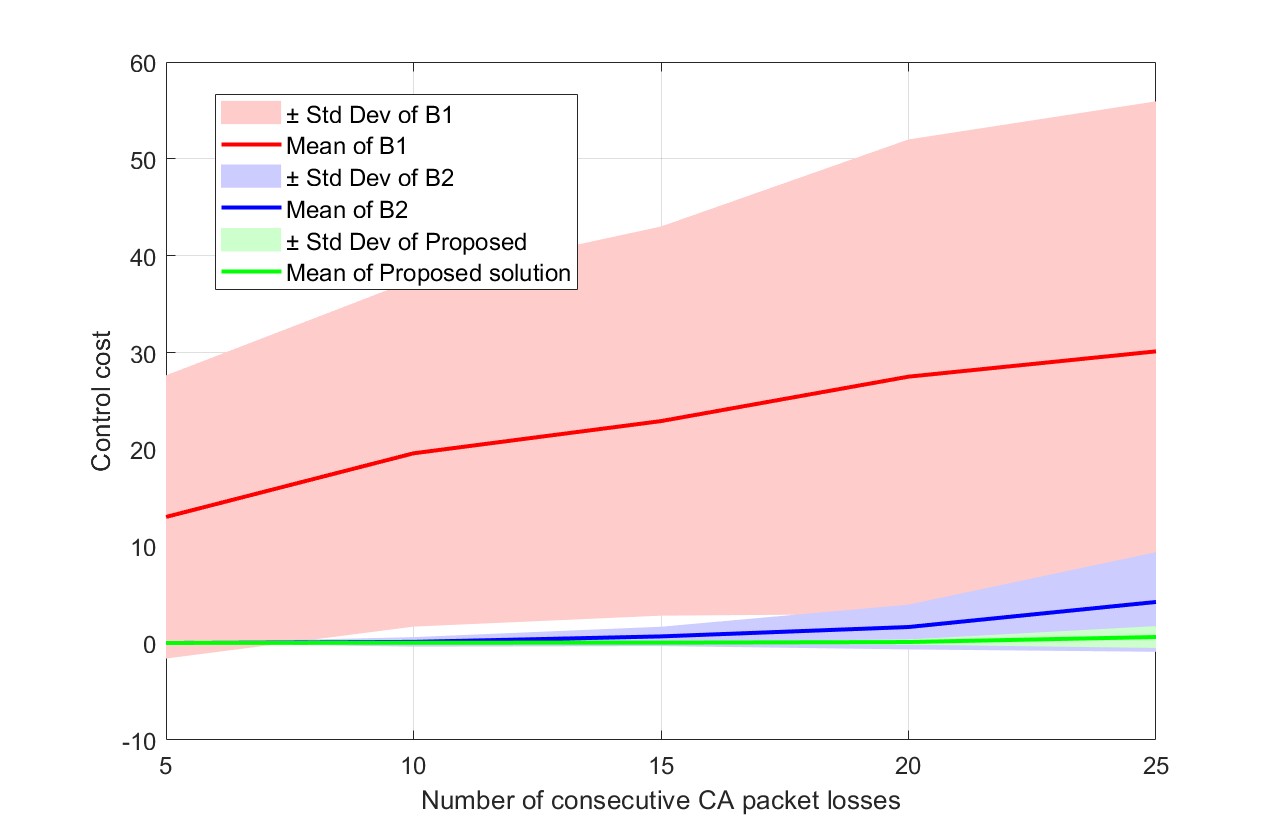}
    \caption{Mean and standard deviation of control cost in~\eqref{eq:our_stabilitycost} over consecutive \gls{ca} link failures, comparing the proposed model with the baselines.}
    \label{fig:consecutive_ca_failure}
\end{figure}

To evaluate the benefits of planning control actions for future consecutive timesteps, we analyzed the control cost ($\frac{1}{T}\sum_{t=0}^{T} J_t$) under scenarios involving consecutive \gls{ca} link failures using three approaches: the proposed method, and two baseline methods B1 and B2.
In this experiment, we considered $N_c = 25$, $T = 200$ and results are averaged over $100$ simulations.
The objective was to assess how well each method maintains control cost in the scenarios of increasing communication failures in \gls{ca} link.

As shown in Fig.~\ref{fig:consecutive_ca_failure}, the control cost increases significantly for B1 as the number of consecutive \gls{ca} failures increases. 
This is expected, as B1 does not apply control action when a \gls{ca} link failure occurs, leading the system to drift further from its stability point. 
B2 method, which reuses the last received control action during \gls{ca} link failures, performs better than B1.
However, as the number of consecutive failures increases ($20$ to $25$), its control cost also rises, and the system's performance deteriorates. 
This suggests that using outdated control actions is not sufficient to maintain stability under large number of consecutive \gls{ca} link failures.
In contrast, the proposed method demonstrates consistently low control costs across all levels of consecutive \gls{ca} link failures. 
The narrow standard deviation indicates stable and reliable performance. 
This confirms that by planning control actions over future time steps, the system can effectively mitigate the impact of communication failures and maintain stability.
In conclusion, the results clearly show that the proposed approach significantly outperforms both baseline methods in scenarios with consecutive \gls{ca} link failures.

\subsubsection{\gls{sc} transmissions with Time}

\begin{figure}
    \centering
    \includegraphics[width=1.1\linewidth]{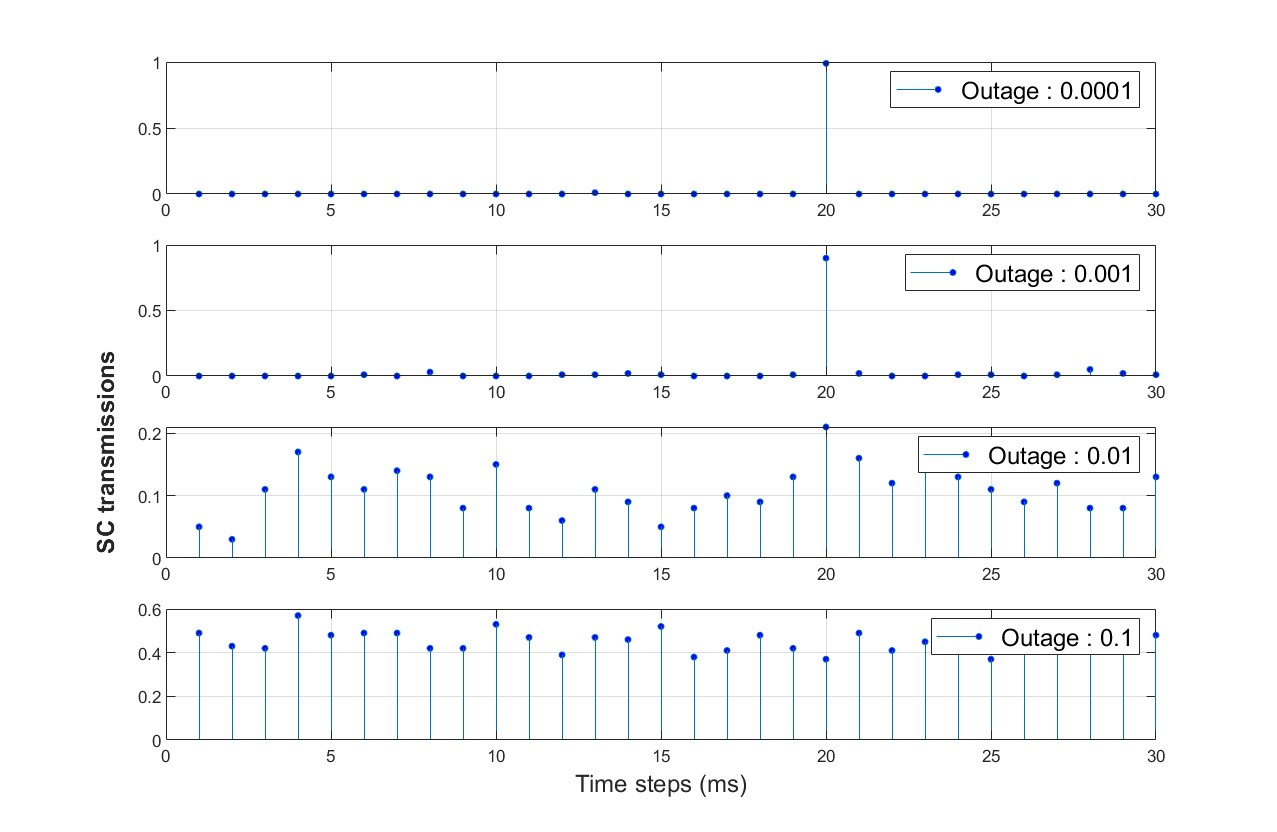}
    \caption{Number of \gls{sc} transmissions over $30$ timesteps for different outage conditions.}
    \label{fig:a_vs_time}
\end{figure}

\begin{table}
    \caption{Statistics of \gls{aoi} and \gls{sc} transmissions with different outage probability}
    \centering
    \begin{tabular}{|c|c|c|c|}
    \hline
        Outage  & Mean & Variance & Number of \gls{sc} \\
        probability & of \gls{aoi} & of \gls{aoi} & transmissions \\[0.5ex]
    \hline
        0.0001 & 19.89 & 1.37 & 50.07 \\
    \hline
        0.001 & 18.53 & 17.39 & 53.45 \\
    \hline
        0.01 & 9.6 & 44.56 & 103.37\\
    \hline
        0.1 & 2.19 & 2.59 & 454.38 \\
    \hline
    \end{tabular}
    \label{table:stat_no_of_a}
\end{table}

The Fig.~\ref{fig:a_vs_time} illustrates the behavior of \gls{sc} link transmissions over time for different outage probability thresholds $(10^{-4}$ to $10^{-1})$ in double pendulum control system with $h'(.) = \tanh(.)$ control. 
The statistical details of \gls{aoi} and the number of \gls{sc}transmissions are provided in Table~\ref{table:stat_no_of_a}. 
In all cases, the \gls{snr} threshold is fixed at $20$\,dBm and $V = 20$.
Moreover, all results shown here and subsequently are based on $T = 1000$ timesteps and averaged over $100$ simulations.
When the outage probability is very low (e.g., $10^{-4}$ or $10^{-3}$), \gls{sc} transmissions are scheduled only after approximately every $20$ consecutive time steps, with no transmissions occurring during the intermediate steps.
This leads to a relatively small total number of \gls{sc} transmissions (around $50$ to $53$) but a large mean \gls{aoi} of approximately $18$ to $20$.
As the outage probability threshold increases, \gls{sc} link transmissions occur more frequently, eventually happening at nearly every time step. 
At the highest outage probability threshold ($10^{-1}$), the system shows the highest number of transmissions $454$ and the lowest mean \gls{aoi} $2.19$, with minimal variance $2.59$. 
This trend arises because a higher outage probability tolerance allows the system to be less strict about potential transmission failures, thus permitting more frequent transmission opportunities.
Therefore, the results highlight the trade-off between communication reliability and transmission frequency: a lower outage probability ensures higher reliability but results in fewer transmission opportunities. In contrast, a higher outage probability increases the risk of transmission failure but enables more continuous and aggressive use of the \gls{sc} link.

\subsubsection{\gls{sc} transmission power usage vs \gls{snr}}

\begin{figure}
    \centering
    \includegraphics[width=1\linewidth]{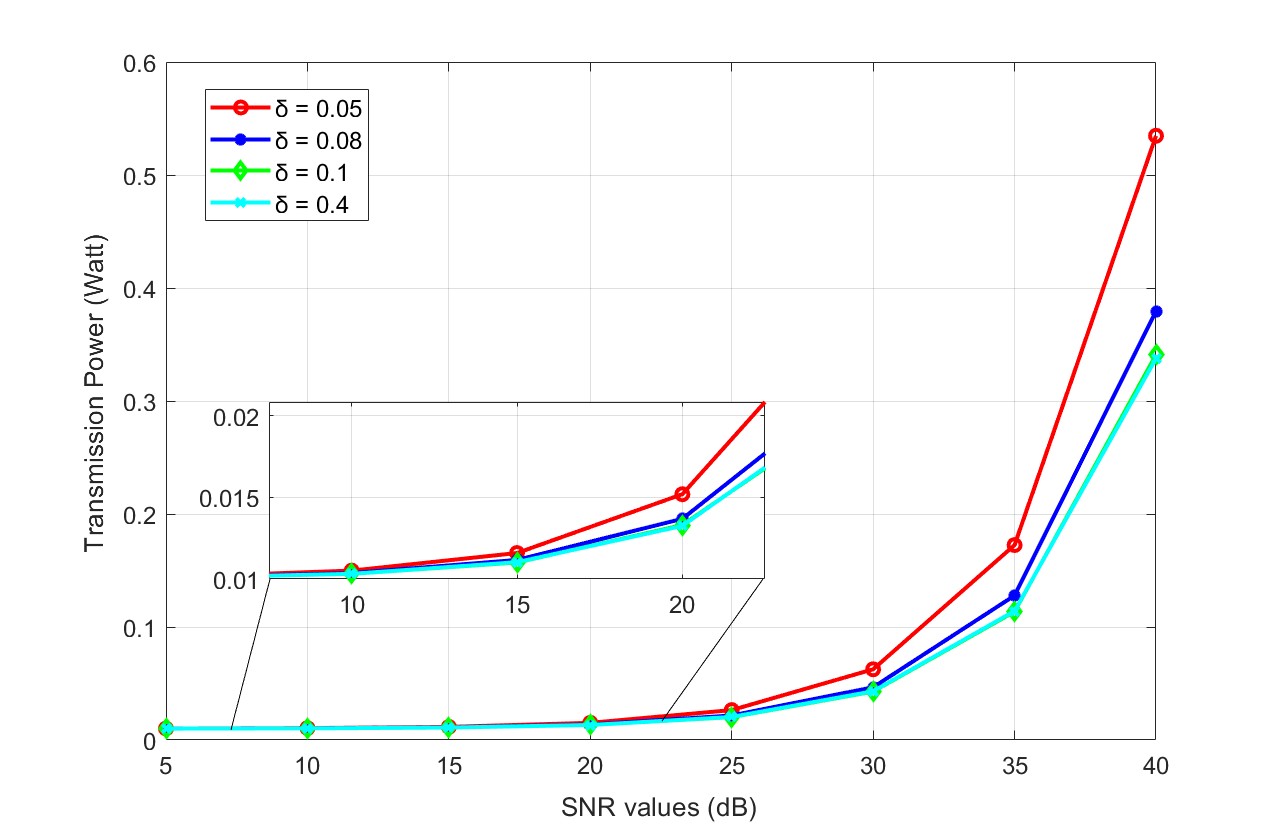}
    \caption{Battery power usage for \gls{sc} Transmission with different \gls{snr} with different error thresholds with $10^{-3}$ outage probability.}
    \label{fig:battery_usage_snr}
\end{figure}

The Fig.~\ref{fig:battery_usage_snr} shows the relationship between the \gls{sc} transmission power consumption and the \gls{snr} under the outage probability threshold of $10^{-3}$ with different $\delta$ values $(0.05, 0.08, 0.1, 0.4)$ for double pendulum setting with the control action ${u} - \frac{1}{3}{u}^3$.
Battery power is consumed for both sensing and \gls{sc} link transmissions, while total sensing power is a constant value in each case.
As the \gls{snr} increases from $5$~dB to $40$\,dB, the \gls{sc} transmission power initially remains very low and nearly constant. 
Specifically, up to around $20$~dB, the power requirement is minimal and shows only slight growth. 
However, beyond $25$~dB, the power consumption begins to increase more noticeably in each scenario due to the increasing demand of the \gls{snr}.

%\subsubsection{Number of planned control actions vs stability}

\subsubsection{Rician factor and SNR with Transmission power}

\begin{figure}
    \centering
    \includegraphics[width=1\linewidth]{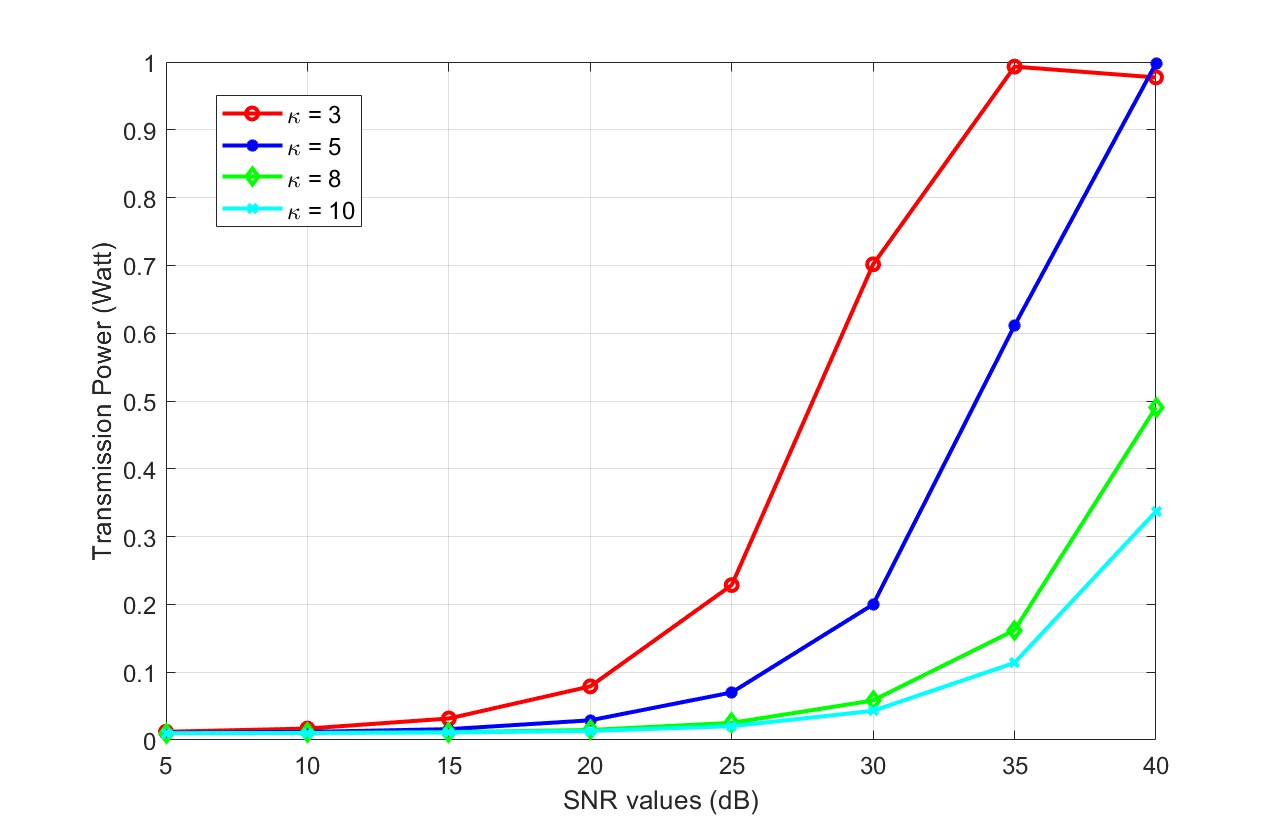}
    \caption{Battery power usage for \gls{sc} Transmission with different \gls{snr} with different rician factor with $10^{-3}$ outage probability.}
    \label{fig:battery_usage_k}
\end{figure}

We have analyzed the impact of the rician factor $\kappa$ on \gls{sc} transmission power consumption and the total cost in~\eqref{eq:mainProblem}.
To this end, we conducted simulations over $1000$ timesteps for different values of $\kappa$ and $\gamma_0$, while maintaining a fixed outage probability threshold $\Tilde{O}$ of $0.001$ in double pendulum setting with ${u} - \frac{1}{3}{u}^3$ control action. 

As shown in Fig.~\ref{fig:battery_usage_k}, the required battery power for \gls{sc} transmission increases with the \gls{snr} threshold, which is expected since higher $\gamma_0$ necessitate stronger signal transmission to meet the outage constraint.
Furthermore, it is evident that lower rician factors such as $\kappa = 3$ require significantly more transmission power than those with higher rician-factors such as $\kappa = 10$ in each \gls{snr}.
This is because a lower rician factor corresponds to a weaker \gls{los} component relative to the scattered components in the wireless channel, leading to more severe fading conditions. 
As a result, more power is required to ensure reliable communication. 
Conversely, higher rician-factors indicate stronger \gls{los} components and more stable channels, allowing for lower power consumption to achieve the same reliability.

\begin{figure}
    \centering
    \includegraphics[width=1\linewidth]{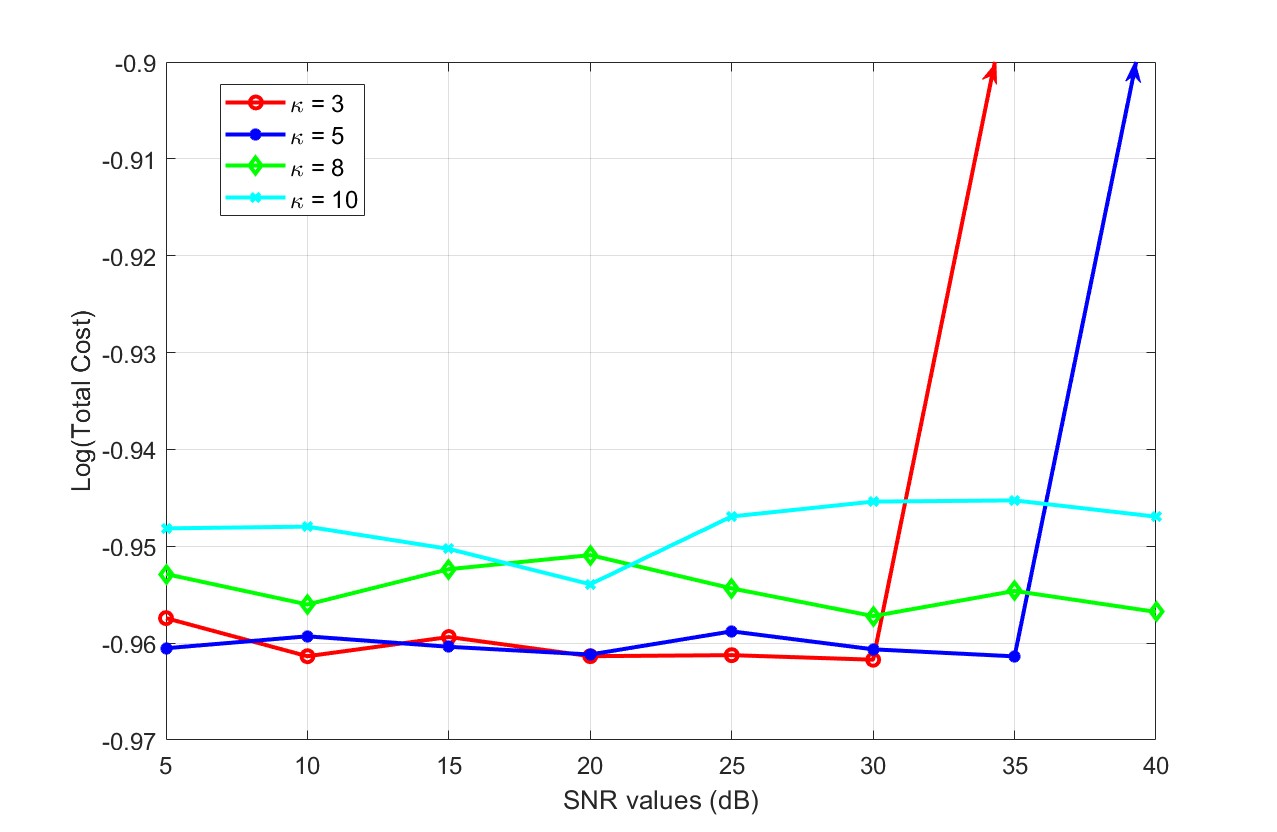}
    \caption{Log value of total cost in~\eqref{eq:mainProblem} with different \gls{snr} with different rician factor with $10^{-3}$ outage probability.}
    \label{fig:totalcost_k}
\end{figure}

Furthermore, Fig.~\ref{fig:totalcost_k} illustrates the log values of total cost, across varying \gls{snr} thresholds for different rician factors $\kappa$. 
As observed, the total cost remains relatively flat at lower \gls{snr} values (e.g. $5$ to $25$) for all $\kappa$.
However, for lower rician factors, such as $\kappa = 3$ and $\kappa = 5$, the total cost tends to increase as the \gls{snr} increases ($\gamma_0 = 35$ and $\gamma_0 = 40$).
This behavior occurs because channels with low rician factors exhibit more severe fading, which necessitates higher transmission power to meet the outage constraint. 
As a result, the available battery power is quickly depleted after a few \gls{sc} transmissions.
As shown in Fig.~\ref{fig:battery_usage_k}, the transmission power reaches up to $1$\,W, which corresponds to the maximum battery power in our scenario.
Once the battery is exhausted, \gls{sc} transmissions cease, preventing the system from providing correct control action.
This leads to deviations from the desired system state and higher control cost $J$, ultimately, an increase in the total cost.
In contrast, for higher rician factors such as $\kappa = 10$, the channel is more stable due to a stronger \gls{los} component, requiring less transmission power to meet outage probability threshold. 
This not only reduces the power required per transmission but also ensures the system can operate over longer time horizons without breaching battery power constraints, thereby maintaining system stability and meeting the objective function.

\section{Conclusion}\label{sec:6}
In this research, we propose a solution that consists of a Deep Koopman operator model for planning control actions and predicting system states in nonlinear, control-non-affine \gls{wncs} and a scheduling algorithm to minimize number of \gls{sc} transmissions in order to optimize limited wireless communication resource usage. 
Simulation results indicate that the proposed model can ensure the stability of nonlinear, control-non-affine \gls{wncs} while requiring fewer transmissions, thereby outperforming existing approaches.
Leveraging this Deep Koopman model, extending this research to multi-agent \gls{wncs} presents an interesting and promising direction for future work.

% --------------- WORKS CITED (10pt FONT) ---------------------
\bibliographystyle{IEEEtran}
%\bibliography{reference}
%\printbibliography

% -------------------------------------------------------------

% -------------------------------------------------------------
\end{document}